\documentclass[aps,twocolumn,showpacs,prb, superscriptaddress,floatfix,longbibliography]{revtex4-2}

\usepackage{amssymb,amsmath,amsthm,bbm,bbold,booktabs,upgreek,ragged2e,multirow,tcolorbox,enumitem}
\usepackage{placeins}
\usepackage{amsfonts}
\usepackage{graphicx}
\usepackage{mathtools}
\usepackage{natbib,siunitx}
\usepackage[breaklinks,colorlinks,allcolors=blue]{hyperref}
\usepackage[version=4]{mhchem}
\usepackage[normalem]{ulem}
\usepackage[dvipsnames]{xcolor}
\usepackage{subfigure}
\usepackage{soul}

\usepackage{pdfpages}
\makeatletter
\AtBeginDocument{\let\LS@rot\@undefined}
\makeatother

\newcommand{\rme}{\mathrm{e}}
\newcommand{\rmd}{\mathrm{d}}

\newcommand{\g}{\gamma}

\newcommand{\G}{\Gamma}
\newcommand{\Gmf}{\G_\mathrm{mf}}
\newcommand{\bra}[1]{\mbox{$\langle #1 |$}}
\newcommand{\ket}[1]{\mbox{$| #1 \rangle$}}
\newcommand{\F}{\mathcal{F}}
\newcommand{\Fc}{\mathcal{F}_\mathrm{c}}
\newcommand{\Fmf}{\mathcal{F}_\mathrm{mf}}
\newcommand{\Ecum}{E_{\mathrm{cum}}}
\newcommand{\ggs}{\g_{\mathrm{gs}}}
\newcommand{\Ggs}{\G_{\mathrm{gs}}}
\newcommand{\Sph}{S_\mathrm{ph}}
\newcommand{\ngsi}{n_{\mathrm{gs},i}}
\newcommand{\rr}{\mathbf{r}}
\newcommand{\iDMFT}{\text{i-DMFT}} 
\DeclareMathOperator{\trace}{Tr}

\begin{document}

\author{Paul G. Graf}
\altaffiliation{These authors contributed equally to this work.}
\affiliation{Arnold Sommerfeld Centre for Theoretical Physics, Ludwig-Maximilians-Universität München, Munich, Germany}

\author{Florian Matz}
\altaffiliation{These authors contributed equally to this work.}
\affiliation{Arnold Sommerfeld Centre for Theoretical Physics, Ludwig-Maximilians-Universität München, Munich, Germany}

\author{Lexin Ding}
\altaffiliation{Present address: ETH Z\"urich, Department of Chemistry and Applied Biosciences, Vladimir-Prelog-Weg 2, CH-8093 Z\"urich, Switzerland}
\affiliation{Arnold Sommerfeld Centre for Theoretical Physics, Ludwig-Maximilians-Universität München, Munich, Germany}

\author{Julia Liebert}
\altaffiliation{Present address: Department of Chemistry, Princeton University, Princeton, New Jersey 08544, USA}
\affiliation{Arnold Sommerfeld Centre for Theoretical Physics, Ludwig-Maximilians-Universität München, Munich, Germany}

\author{Markus Penz}
\affiliation{Arnold Sommerfeld Centre for Theoretical Physics, Ludwig-Maximilians-Universität München, Munich, Germany}
\affiliation{Department of Computer Science, Oslo Metropolitan University, Norway}

\author{Christian Schilling}
\affiliation{Arnold Sommerfeld Centre for Theoretical Physics, Ludwig-Maximilians-Universität München, Munich, Germany}
\email{c.schilling@lmu.de}


\title{Capturing electron correlation at mean-field cost:\\
Assessment of i-DMFT and the underlying correlation conjecture}

\date{\today}

\begin{abstract}
Accurately treating strong electron correlation in quantum chemistry typically requires multireference wave-function methods with steep computational scaling. The recently proposed i-DMFT method promises near configuration-interaction accuracy at mean-field cost by invoking an empirical linear relation between correlation energy and entropy (Collins' conjecture), whose validity remains unclear. We systematically assess this relation across a range of di- and polyatomic molecules, including diverse bond types, third-row elements, different types of geometric distortions, and excited states. We find that the conjectured linearity holds for bond-breaking processes dominated by electron redistribution within orbital pairs, but breaks down for heterolytic dissociation and excited states. In simple molecules, i-DMFT provides a reasonable description of total energies, but does not reliably reproduce reduced density matrices or individual energy components. It further degrades in more complex cases such as ethylene. Based on these results, we formulate criteria for the validity of the conjecture and outline implications for entropy-based reduced density matrix functionals.
\end{abstract}

\maketitle

\section{Introduction}\label{sec:intro}

Computational modeling of the structure, reactivity, and properties of molecular systems and solids has become an indispensable part of modern research in chemistry and material science. However, there remains a pressing need for computationally efficient and accurate approaches capable of reliably describing the complex systems encountered in frontier research, which often involve strong correlation, near-degeneracies, and low-lying excited states~\cite{zhou22}. Achieving high accuracy at manageable computational cost is therefore a central challenge, and further progress will likely require fundamentally new theoretical ideas.

Among existing approaches, many wave-function methods such as coupled-cluster theory~\cite{bartlett11,verma20} rely on a single-reference description and therefore struggle with strong static correlation~\cite{coe15}. More advanced methods, including multireference coupled-cluster theory~\cite{koehn12} and complete active space perturbation theory~\cite{andersson92}, overcome this limitation by employing multiconfigurational references and incorporating dynamic correlation on top of these. Their improved accuracy, however, comes at a strongly increased computational cost. Other methods aim to approach full configuration interaction (FCI) accuracy through systematically improvable or adaptively truncated expansions of the configuration space, for instance via the density-matrix renormalization group~\cite{verstraete23} or selected CI methods~\cite{holmes16,liu16,park21}.

Density-functional theory (DFT), the workhorse of modern computational chemistry, offers a favorable balance between accuracy and efficiency, but standard Kohn-Sham (KS) implementations struggle to describe strong static correlation~\cite{yu16}, for example in bond-breaking situations.
Extensions beyond single-determinant KS-DFT, including spin-adapted~\cite{filatov99} and multiconfigurational approaches~\cite{sharma21}, aim to address these limitations, yet a universally reliable and efficient description of strongly correlated systems remains elusive.

In this context, reduced-density-matrix functional theory (RDMFT)~\cite{Gilbert75, DP78, Levy79, Valone80, M84, Goedecker1998, GPB05, SDLG08, Pastor2011a, SDSG13, cioslowski2015a, PG16, rodriguez-mayorga2017a, SKB17, GUL18, Schilling18, vMGB18, GR19, Cioslowski19, CS19ExForce, SP21, DSKBR22, Mazziotti22, LCLS21, LKO22, SYNF22, liebert23, LS23-njp, LS23-sp, CG24, Piris24-review, VMSS24, YS24, Fredheim2025, HLYMdCP25} has attracted increasing attention due to its potential to describe electron correlation through fractional occupation numbers without explicit access to the full many-body wave function. Static correlation is reflected in the distribution of electrons over multiple orbitals, while dynamic correlation is captured by the tail of the occupation spectrum in sufficiently large basis sets~\cite{Giesbertz2013,benavidesriveros17}. Unlike DFT, however, RDMFT lacks a non-interacting reference system at zero temperature, which generally prevents mean-field computational scaling.

The recently proposed i-DMFT method~\cite{WangBaerends22-PRL} builds on this framework by introducing a simple approximate functional that enables calculations at mean-field cost, comparable to the Hartree-Fock method. It is based on a conjectured linear relation between the correlation energy and the entropy of the one-particle reduced density matrix (1RDM), known as Collins' conjecture. Related “correlation conjectures” have appeared in various forms~\cite{Ziesche1995,WangWangSheng2021,WangKnowles21-PRA}, all aiming to provide a simple route to the correlation energy and rooted in information-theoretic concepts.

Despite the absence of a rigorous theoretical foundation and earlier critical analysis~\cite{Cioslowski2024-CC-Fallacy}, empirical evidence for such linear relations has been reported for several systems, primarily diatomic molecules along dissociation coordinates~\cite{cioslowski23,WangWangSheng2021,WangKnowles21-PRA,WangBaerends22-PRL,Irimia2023-open-shell,Hu2024-dispersion,Liu2024-halogen,Liu2025-halides,Irimia2023-correlation}, {and a formal justification for their use was given~\cite{MartinezB2023}}. However, systematic benchmarks of the conjecture and its implications for functional approximations remain scarce. In this work, we perform a comprehensive assessment across diatomic and polyatomic molecules, including geometric deformations and excited states, and thereby identify the regimes in which the conjectured linearity holds and those in which it breaks down. Based on this analysis, we derive practical criteria for the validity of the conjecture and critically evaluate the performance and limitations of i-DMFT.

This paper is structured as follows: Sec.~\ref{sec:theory} reviews the theoretical foundations of RDMFT and i-DMFT. In Sec.~\ref{sec:assess-cc}, we assess the validity of the conjecture using high-level wave-function methods across diatomics, polyatomic systems, geometric deformations, and excited states. The performance of i-DMFT is then evaluated in Sec.~\ref{sec:assess-iDMFT} in terms of energy surfaces, natural occupation numbers, reduced density matrices, electron densities, and individual energy components. Finally, Sec.~\ref{sec:concl} summarizes the findings, formulates validity criteria of the CC, and outlines perspectives for entropy-based density matrix functionals.

\section{RDMFT and \MakeLowercase{i}-DMFT Basics}
\label{sec:theory}

In this section, we briefly summarize the basic formalism of RDMFT~\cite{Gilbert75,Levy79,Valone80,PG16,liebert23} and introduce i-DMFT as a special variant.

RDMFT is a method for solving the ground-state problem for Hamiltonians $\hat H = \hat h + \hat W$, where $\hat h$ denotes the one-body contribution and $\hat W$ denotes a general two-body term. In the context of electronic structure theory, which is the focus of this work, $\hat W$ corresponds to the Coulomb interaction. While $\hat W$ is considered fixed, the set of $\hat h$ formally consists of the space of all self-adjoint one-particle operators, thereby defining the scope of the theory~\cite{liebert23,UnifiedFT2026-preprint}.

As the central object of the theory, the universal functional $\F(\g)$ of an $N$-particle system is defined by a constrained search over all $N$-representable two-particle reduced density matrices (2RDMs) $\Gamma$ that yield a given one-particle reduced density matrix (1RDM) $\g$ (denoted as ``$\G \mapsto \g$''),
\begin{equation}\label{eq:F-def}
    \F(\g) = \inf_{\Gamma \mapsto \g} \trace[\hat W \Gamma].
\end{equation}
The functional is \emph{universal} in the sense that it depends only on the interaction $\hat W$ and the particle number $N$, but not on the one-particle Hamiltonian $\hat h$, and is therefore applicable to a wide range of chemical systems. The restriction of the constrained search to $N$-representable 2RDMs, rather than the full many-body state, is possible because $\hat W$ is a two-body operator.

For any fixed choice of $\hat h$, minimization of the total-energy functional
\begin{equation}
    E(\g) = \trace[\hat h \g] + \F(\g)
\end{equation}
yields the ground-state energy and the corresponding 1RDM $\ggs$ of $\hat H = \hat h + \hat W$. The method is, in principle exact, provided that the exact functional $\F(\g)$ is known.

It is often convenient to express $\g$ (a self-adjoint operator on the one-particle Hilbert space) in its spectral representation,
\begin{equation}\label{1RDM}
\g = \sum_i n_i \ket{\chi_i}\bra{\chi_i}.
\end{equation}
The eigenvalues $\{n_i\}$ are referred to as \emph{natural occupation numbers} (NONs), and the corresponding eigenvectors $\{\chi_i\}$ as \emph{natural orbitals} (NOs).

A zeroth-level approximation to $\F(\g)$ is its mean-field contribution,
\begin{eqnarray}\label{eq:HF}
  \Fmf(\g) &=& \trace[\hat W \Gmf(\g)] \nonumber \\
           &=& \frac{1}{2} \sum_{i,j} n_i n_j \bra{\chi_i \chi_j} \hat W \ket{\chi_i \chi_j},
\end{eqnarray}
obtained by evaluating $\trace[\hat W \Gamma]$ in Eq.~\eqref{eq:F-def} at
$\Gmf(\g) = \frac{1}{2}(\hat I - \hat X)\,\g \otimes \g$,
with $\hat X(\ket{\chi_i}\otimes\ket{\chi_j})=\ket{\chi_j}\otimes\ket{\chi_i}$ and
$\ket{\chi_i \chi_j} = (\ket{\chi_i}\otimes\ket{\chi_j} - \ket{\chi_j}\otimes\ket{\chi_i})/\sqrt{2}$.

Minimizing $\trace[\hat h \g] + \Fmf(\g)$ over all 1RDMs yields the Hartree-Fock solution~\cite{LiebHF, Lieb1977}, corresponding to an idempotent 1RDM, $\g^2 = \g$, i.e., $n_i \in \{0,1\}$. The deviation from the exact functional defines the correlation functional,
\begin{equation}
    \Fc(\g) = \F(\g) - \Fmf(\g) \leq 0,
\end{equation}
where the inequality holds if the interaction $\hat W$ is positive semidefinite~\cite{LiebHF,Levy87}.
This decomposition corresponds to the cumulant expansion of the two-particle reduced density matrix~\cite{Kutz99} and differs fundamentally from defining correlation energy as the difference to the Hartree-Fock energy~\cite{Lowdin1955}.

At the level of quantum states, or equivalently at the level of their associated two-particle reduced density matrices, one instead considers the \emph{cumulant energy},
\begin{equation}\label{eq:Ecum}
    \Ecum(\Gamma) = \trace[\hat W(\Gamma - \Gmf(\g_\Gamma))],
\end{equation}
where $\g_\Gamma$ denotes the 1RDM associated with $\Gamma$. The correlation functional can then be written as
\begin{equation}\label{eq:Fc-def}
    \Fc(\g) = \inf_{\Gamma \mapsto \g} \Ecum(\Gamma).
\end{equation}
Since the ground-state energy is obtained by minimizing $\trace[\hat h \gamma] + \Fmf(\gamma) + \Fc(\gamma)$, a ground state with $\Ggs \mapsto \ggs$ must also realize the minimum in Eq.~\eqref{eq:Fc-def} (which is known to exist~\cite{Fredheim2025}),
\begin{equation}
    \Fc(\ggs) = \Ecum(\Ggs).
\end{equation}
{We note here that $\Ecum$ is also referred to as intrinsic correlation energy \cite{kutzelnigg2003}. In order to emphasize distinctness from the correlation energy, which appears in Collins' original conjecture \cite{Collins}, we chose to use the term ``cumulant energy'' for $\Ecum$.}

In order to make RDMFT practical, one must construct approximations to the correlation functional that are accurate within a given domain of interest. In this context, \citet{WangBaerends22-PRL} proposed i-DMFT, which is based on the empirical observation of an approximately linear relation between the cumulant energies of ground states and the particle-hole symmetrized entropies associated with their one-particle reduced density matrix. This relation has been observed along dissociation coordinates of certain molecular systems and is referred to as the correlation conjecture (CC),
\begin{equation}\label{eq:Collins}
\Ecum(\Ggs) \approx -\kappa \Sph(\{\ngsi\}) - b.
\end{equation}
Here, $\kappa$ and $b$ are the system-dependent parameters, and $\Sph$ denotes the particle-hole symmetrized von Neumann entropy of a 1RDM, which reduces to 
\begin{equation}\label{eq:Sph}\begin{aligned}
\Sph(\{n_i\}) = -\sum_{i} [&n_i \ln(n_i)\\
&+(1-n_i) \ln(1-n_i)].
\end{aligned}\end{equation}
In Eq.~\eqref{eq:Collins}, $\Sph$ is evaluated for the NONs $\{\ngsi\}$ of the ground-state 1RDM $\ggs$. We emphasize that $\Sph$ is not the entropy of the many-body state, but a quantity derived solely from the spectrum of the 1RDM. {We note that $\Sph$ amounts exactly to the nonfreeness of the underlying many-body state that is well known in quantum information theory~\cite{Gottlieb2007Properties}.}

This observation motivates approximating the correlation functional as
\begin{equation}\label{eq:Fc-approx}
    \Fc(\g)\approx -\kappa\Sph(\{n_i\}) - b.
\end{equation}
By expanding the logarithm, this form can also be related to a 1RDM correction to DFT~\cite{Gibney2022}. However, the validity of Eq.~\eqref{eq:Collins} cannot be expected for arbitrary states. For any $\G \mapsto \ggs$, the inequality $\Ecum(\G) \geq \Fc(\ggs)$ holds, and together with Eq.~\eqref{eq:Fc-approx} (which assumes the CC) this implies
\begin{equation}\label{eq:Ecum-inequality}
    \Ecum(\G) \geq -\kappa\Sph(\{\ngsi\}) - b.
\end{equation}
If this condition is violated, the i-DMFT functional cannot select the correct ground state.

The resulting i-DMFT approximation~\cite{WangBaerends22-PRL} reads
\begin{equation}\label{func}
\F^\iDMFT(\g) = \Fmf(\g) - \kappa\Sph(\{n_i\}) - b,
\end{equation}
where $\{n_i\}$ are the NONs of $\g$. The variation of the total-energy functional
\begin{equation}
    E(\g) = \trace[\hat h \g] + \F^\iDMFT(\g)
\end{equation}
with respect to the NOs yields the single-particle equations~\cite{WangBaerends22-PRL}
\begin{equation}\label{eq:th-HF}
    \left(\hat h + \sum_j n_j(\hat J_j - \hat K_j)\right) \chi_i = \varepsilon_i \chi_i,
\end{equation}
where $\hat J_j$ and $\hat K_j$ are the Hartree and exchange operators, respectively, as in Hartree-Fock theory. Variation with respect to the NONs yields the Fermi-Dirac distribution
\begin{equation}\label{eq:FD-dist}
    n_i = \left(1 + \rme^{(\varepsilon_i - \mu)/\kappa}\right)^{-1},
\end{equation}
where the chemical potential of the electrons $\mu$ can be determined from the normalization condition $\sum_in_i=N$. Together, Eqs.~\eqref{eq:th-HF} and \eqref{eq:FD-dist} can be identified exactly as the Hartree-Fock method with an electronic temperature $\kappa/k_\mathrm{B}$~\cite{Mermin65}. Because of its conceptual similarities to the Hartree-Fock method, i-DMFT can be easily implemented and tested, which is the topic of Sec.~\ref{sec:assess-iDMFT}.

{Even though the Fermi-Dirac distribution is applicable in many physical settings, recent works in the framework of RDMFT~\cite{CS19ExForce,Cioslowski2024-CC-Fallacy} have revealed the correct behavior of the natural occupation numbers in quantum chemical systems to deviate significantly from Eq.~\eqref{eq:FD-dist}. The consequences of this for the quality of the predictions made via i-DMFT will become evident in Sec.~\ref{sec:assess-iDMFT}.}

There are a few aspects of i-DMFT that can already be discussed from a theoretical perspective before we move on to a numerical assessment.
First, the need to determine $\kappa$ and $b$ for different molecules conflicts with the universality principle of RDMFT, where the functional $\Fc(\gamma)$ is, in principle, the same for all systems with a given interaction $\hat{W}$. However, this does not necessarily limit the usefulness of the method, since it may still have a well-defined range of applicability. In addition, although the computational cost of i-DMFT is comparable to that of Hartree-Fock or KS-DFT, the parameters $\kappa$ and $b$ must be provided as input and typically require high-level calculations.

From a mathematical point of view, functional theories are usually based on a strict \emph{duality principle}~\cite{Lieb83,engel2011,Penz2023-ReviewPartI,liebert23}. This means that there is a one-to-one correspondence, in the sense of vector spaces, between potentials (which define the chemical system) and internal quantities that describe the state. In DFT, this correspondence is between the scalar potential and the one-particle density, while in RDMFT it is between the one-particle Hamiltonian $\hat h$ and the 1RDM $\gamma$. In i-DMFT, however, the correlation functional depends only on a single scalar quantity, namely the entropy, whereas the parameter space of molecular deformations can involve many more degrees of freedom. This raises the question: which deformations in a given molecule are covered by one single pair $(\kappa,b)$? It is worth noting that many examples in the literature, as well as in this work, focus on the dissociation of diatomic molecules, which involves only a single degree of freedom. Nevertheless, trial calculations on H$_2$O~\cite{WangKnowles21-PRA} suggest that the CC may also hold in higher-dimensional parameter spaces. The limited discussion of more complex geometries, together with our goal of broader chemical applicability, motivated us to consider such cases in this work.

The fact that only a single parameter $\kappa$ couples to an internal quantity, here the entropy $\Sph$, suggests that the method may not capture more complex properties of the system.
By construction, i-DMFT mainly targets quantities such as the ground-state energy and the entropy entering the CC, and indirectly the NON spectrum, which determines the entropy in Eq.~\eqref{eq:Sph}. In contrast, the mechanism provided to reproduce the full 1RDM beyond its eigenvalues, i.e., the natural orbitals, the electron density, and general one- and two-particle observables, is questionable. We will demonstrate these limitations in Sec.~\ref{sec:assess-iDMFT}. As a first step, we assess the validity of the CC itself in Sec.~\ref{sec:assess-cc}.

\section{Assessment of the correlation conjecture \label{sec:assess-cc}}

An assessment of the validity of the CC in the context of chemical applications ultimately requires evaluating cumulant energy and entropy for exact eigenfunctions of the physical Hamiltonian, which are notoriously difficult to obtain. Our analysis is based on CASSCF (complete active space self-consistent field) calculations, aiming at convergence with respect to both basis set and active space size.

We select molecules that represent a diverse set of common chemical bonds and analyze the relation between cumulant energy and particle-hole symmetrized entropy along different geometric deformations. Where approximately linear behavior consistent with the CC is observed, we estimate $\kappa$ and $b$ by linear fits that minimize the weighted root mean square error $\bar{\Delta}$ {in the cumulant energies as a function of $\Sph$ along the dissociation curve}. Details of the fitting procedure, as well as dissociation energies $D_\mathrm{e}$, equilibrium geometries ($r_\mathrm{e}$, $\theta_\mathrm{e}$), harmonic frequencies $\tilde{\nu}$, and stiffness parameters $k_2$, are provided in the supplementary material together with the specification of the sampled geometries.

\subsection{Two-electron systems}\label{sec:two_elec_sys}

We begin our analysis with diatomic molecules, which serve as prototypical systems for the study of strong correlation, chemical bonding, and electronic structure. In addition to the availability of accurate reference data~\cite{huber79}, their small number of electrons allows for near-FCI accuracy. Their dissociation provides a paradigmatic example of the gradual emergence of strong correlation, where single-reference methods such as Hartree-Fock or KS-DFT fail to describe the potential energy curves accurately despite the simplicity of the systems.

\begin{figure}
\includegraphics[width=\linewidth]{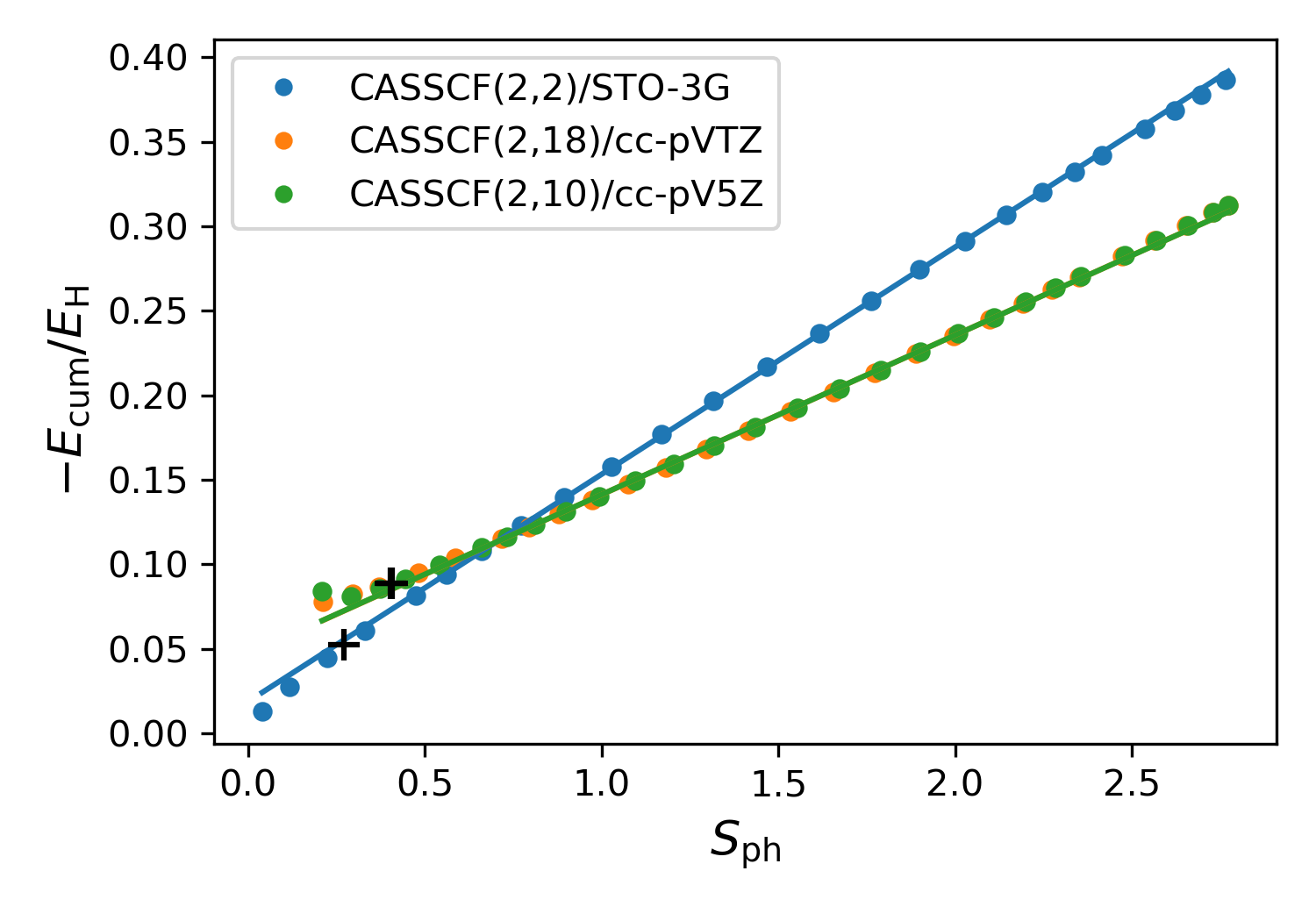}
\caption{Cumulant energy $\Ecum$ vs particle-hole symmetrized entropy $\Sph$ for the dissociation of the hydrogen molecule.}
\label{fig:h2results}
\end{figure}

We first consider two-electron systems, which can be treated with particularly high accuracy. The simplest example is the hydrogen molecule. The results for its dissociation are shown in Table~\ref{tab:h2results} and Fig.~\ref{fig:h2results}.

\begin{table*}
\caption{Fit results from the dissociation curve and the $\Sph$--$\Ecum$ dependence for the hydrogen molecule{, evaluated with CASSCF(2,$n_\mathrm{orb}$) calculations in different basis sets}.}\label{tab:h2results}
\begin{tabular}{lclllllr}\toprule
Basis&${n_\mathrm{orb}}$&$D_\mathrm{e}$/$E_\mathrm{H}$&$r_\mathrm{e}$/bohr&$\tilde{\nu}$/cm$^{-1}$&$\kappa$/$E_\mathrm{H}$&$b$/$E_\mathrm{H}$&$\bar{\Delta}$/m$E_\mathrm{H}$\\\midrule
STO-3G&2&0.204&1.39&5011&0.134&0.0187&2.33\\
cc-pVDZ&2&0.149&1.46&4192&0.105&0.0232&2.55\\
cc-pVDZ&10&0.165&1.44&4370&0.0967&0.0413&1.47\\
aug-cc-pVDZ&18&0.166&1.44&4318&0.0960&0.0418&1.52\\
cc-pVTZ&2&0.171&1.43&4210&0.105&0.0233&1.89\\
cc-pVTZ&10&0.171&1.41&4422&0.0948&0.0458&2.45\\
cc-pVTZ&18&0.172&1.41&4438&0.0944&0.0467&2.70\\
cc-pVQZ&10&0.172&1.41&4407&0.0944&0.0468&2.69\\
cc-pV5Z&10&0.172&1.41&4408&0.0943&0.0468&2.71\\\midrule
Exp.&&0.165~\cite{liu09}&1.40~\cite{huber79}&4401~\cite{huber79}\\\bottomrule
\end{tabular}
\end{table*}

At equilibrium distance, the hydrogen molecule is well described by a single-reference wave function, with an occupation number close to two for the bonding $1\upsigma_g$ orbital, resulting in a small entropy $\Sph$. During dissociation, the energies of the $1\upsigma_g$ and $1\upsigma_u$ orbitals approach each other, and their occupation numbers become equal. In the dissociation limit, the wave function is dominated by two configurations, such that the two lowest, degenerate orbitals are half occupied and $\Sph = 4\ln 2 = 2.773$~\cite{cioslowski23}. As predicted by the CC, the cumulant energy increases approximately linearly with the entropy. Small deviations from linearity occur at short bond lengths, especially when using larger basis sets.

The determined parameters $\kappa$ and $b$ change when moving from STO-3G to larger basis sets, but already the cc-pVDZ basis set provides a reliable estimate of their values in the basis set limit. While $\kappa$ decreases only weakly when increasing the active space beyond two orbitals, $b$ converges only for active spaces of about ten orbitals. This suggests that $\kappa$ is primarily governed by static correlation, while $b$ increases with the presence of dynamic correlation --- an interpretation that will be substantiated by the examples discussed below. This can be rationalized by noting that the cumulant energy combines contributions from both static and dynamic correlation. Dynamic correlation remains significant even for states that are close to single-reference wave functions. Although dynamic correlation also leads to a nonzero entropy, it does so via a different mechanism, namely the so-called ‘fermionic exchange force’~\cite{CS19ExForce}, which induces small deviations from the extremal occupation numbers (0 and 1) and can result in a large correlation energy even at small entropies. {Notably, while the modifications to the chemical systems considered here do indeed weakly affect the magnitude and character of dynamic correlation as can be analyzed by evaluating the natural densitals~\cite{cioslowski25}, this type of correlation still} appears as an approximately constant {contribution} along dissociation {processes}. By contrast, static correlation scales very differently with entropy, as it evolves gradually during bond-breaking processes and is not merely due to small corrections to the occupation numbers. The CC analysis attempts to separate these contributions into a part that varies with $\Sph$ and a constant offset, which formally leads to the unphysical result that $\Ecum$ is nonzero at $\Sph = 0$.

\begin{figure}
\includegraphics[width=\linewidth]{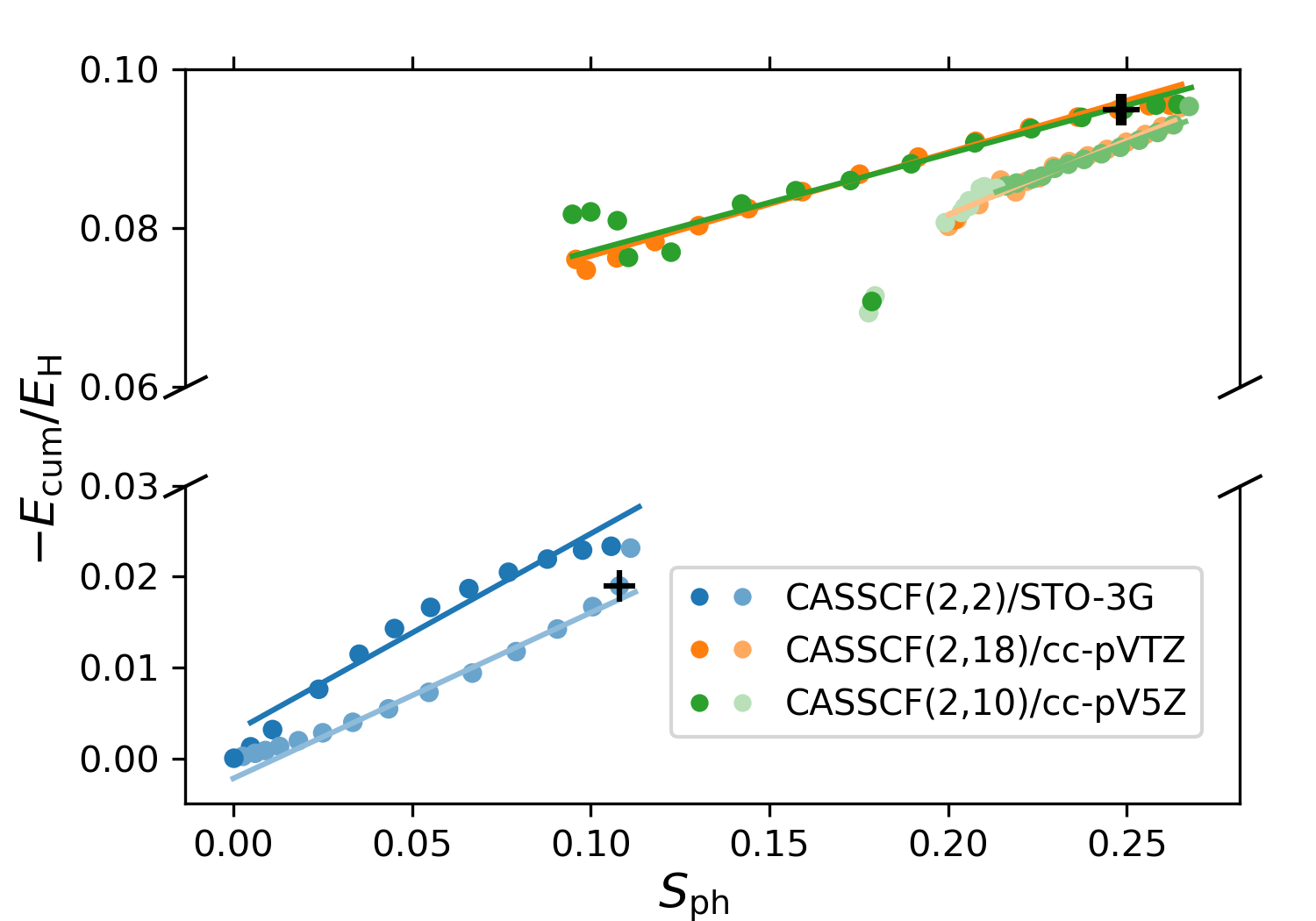}
\caption{Cumulant energy $\Ecum$ vs particle-hole symmetrized entropy $\Sph$ for the dissociation of the helium hydride cation. Dark shades: $R < R_\mathrm{turn}$. Lighter shades: $R\geq R_\mathrm{turn}$. STO-3G: $R_\mathrm{turn}=1.65$~bohr and cc-pV$n$Z: $R_\mathrm{turn}=2.15$~bohr. Very light shades for cc-pV$n$Z: $R\geq 4.43$~bohr, not included in the fitting procedure.}
\label{fig:hehresults}
\end{figure}

Replacing one of the atoms in H$_2$ with helium leads to a qualitatively different electronic structure and $\Ecum$--$\Sph$ relation. The bond strength decreases significantly, and upon dissociation a finite gap remains between the two valence orbitals, such that the electrons stay localized on the helium nucleus~\cite{loreau10}. The bond in this strongly polarized system is largely coordinative, making this example particularly relevant, as such non-covalent interactions are ubiquitous in chemical systems. The results for the dissociation are shown in Table~\ref{tab:hehresults} and Fig.~\ref{fig:hehresults}.

\begin{table*}
\caption{Fit results from the dissociation curve and the $\Sph$--$\Ecum$ dependence for the helium hydride cation{, evaluated with CASSCF(2,$n_\mathrm{orb}$) calculations in different basis sets}.}\label{tab:hehresults}
\begin{tabular}{lclllrllrrr}\toprule
&&&&&\multicolumn{3}{c}{$R$ small}&\multicolumn{3}{c}{$R$ large}\\
Basis&${n_\mathrm{orb}}$&$D_\mathrm{e}$/$E_\mathrm{H}$&$r_\mathrm{e}$/bohr&$\tilde{\nu}$/cm$^{-1}$&$\kappa$/$E_\mathrm{H}$&$b$/$E_\mathrm{H}$&$\bar{\Delta}$/m$E_\mathrm{H}$&$\kappa$/$E_\mathrm{H}$&$b$/$E_\mathrm{H}$&$\bar{\Delta}$/m$E_\mathrm{H}$\\\midrule
STO-3G&2&0.0549&1.72&2742&0.220&0.00281&1.71&0.180&-0.00211&0.889\\
cc-pVDZ&2&0.0898&1.50&3242&0.213&0.0157&1.37&0.200&0.0122&0.291\\
cc-pVDZ&10&0.0733&1.49&3243&0.229&0.0310&1.67&0.162&0.0414&0.385\\
aug-cc-pVDZ&18&0.0722&1.49&3141&0.213&0.0351&1.94&0.172&0.0389&0.298\\
cc-pVTZ&2&0.0925&1.47&3200&0.206&0.0190&1.20&0.187&0.0172&0.284\\
cc-pVTZ&10&0.0765&1.47&3175&0.128&0.0620&0.708&0.162&0.0485&0.414\\
cc-pVTZ&18&0.0774&1.47&3204&0.131&0.0634&0.807&0.188&0.0440&0.530\\
cc-pVQZ&10&0.0764&1.47&3213&0.109&0.0674&0.800&0.167&0.0485&0.441\\
cc-pV5Z&10&0.0764&1.47&3211&0.122&0.0649&1.79&0.165&0.0494&0.202\\\midrule
Exp.&&0.0750~\cite{huber79,pachucki12}&1.46~\cite{bernath81}&2911~\cite{perry14}\\
\bottomrule
\end{tabular}
\end{table*}

The heterolytic character of the bond dissociation leads to the electronic structure remaining single-configurational throughout the entire dissociation: the entropy of the 1RDM stays below 0.3. At small interatomic distances, the $\upsigma$ orbitals exhibit bonding and antibonding character. The occupation of the antibonding orbital is nonzero due to dynamic correlation and initially increases with $R$. Beyond a threshold distance $R_\mathrm{turn}$ (1.65~bohr for CASSCF(2,2)/STO-3G, 2.16~bohr for CASSCF(2,18)/cc-pVTZ, and CASSCF(2,10)/cc-pV5Z), the orbitals start to become localized and the 1s orbital of the hydrogen atom remains unoccupied, such that the occupation of the $2\upsigma$ orbital decreases again, reducing $\Sph$. In the minimal basis set STO-3G, the $\Sph$--$\Ecum$ curve forms a round loop, with both ends approaching $\Sph=\Ecum=0$, which would imply that no part of the dissociation follows the CC. However, in larger basis sets, the energy converges toward the physical limit, where the endpoints become distinct and approximate linearity is recovered in both the small-$R$ and large-$R$ regimes. This suggests that the CC, and thus i-DMFT, may still be reliable in larger basis sets, at least within certain parts of the dissociation process. {Nonetheless, the significant change in electron distribution, associated with the heterolytic bond breaking in this example, leads to a departure from global linearity. A thought experiment on such effects has previously been discussed as an argument against the universality of the CC~\cite{Cioslowski2024-CC-Fallacy}.}

This behavior differs from that of the hydrogen molecule and from most of the examples discussed below, where larger basis sets typically lead to larger deviations from linearity. This can again be attributed to an improved description of dynamic correlation: the underlying assumption of the CC, namely that dynamic correlation is approximately independent of the dissociation coordinate, is not strictly fulfilled.

However, in the helium hydride molecule, dynamic correlation plays an even more important role in describing bond breaking, as it is required to capture the electron density around the helium atom that attracts the hydrogen nucleus. This indicates that even for the dissociation of coordinative bonds, the CC may hold in certain regimes. {Additionally, helium hydride is a prototypical example of a molecule subject to dispersion interactions. An ab initio description of these effects has especially arduous requirements toward electron correlation and a reduced density matrix description requires an identification of the long-range part of the cumulant~\cite{vianadal17,vianadal19}. As stated previously, dynamic correlation effects are included in $b$.} Between cc-pVTZ and cc-pV5Z, $\kappa$ and $b$ still vary, suggesting that larger basis sets or active spaces may be required to reach convergence, in contrast to the hydrogen case. Qualitatively, $\kappa$ is larger for the large-$R$ branch of the curve than for the small-$R$ branch. Compared to the hydrogen molecule, $\kappa$ is smaller by roughly a factor of 3 (large $R$) or 4 (small $R$), while $b$ is similar (large $R$) or larger (small $R$), reflecting the effect of the modified one-body potential.

\subsection{Second-row elements}\label{sec:secondrow}

In this section, we investigate diatomics composed of second-row elements, as they exhibit a wide range of bond strengths and electronic configurations. These molecules are chemically highly relevant and therefore particularly well suited for assessing the CC and its usefulness in computational chemistry.

\begin{figure}
\includegraphics[width=\linewidth]{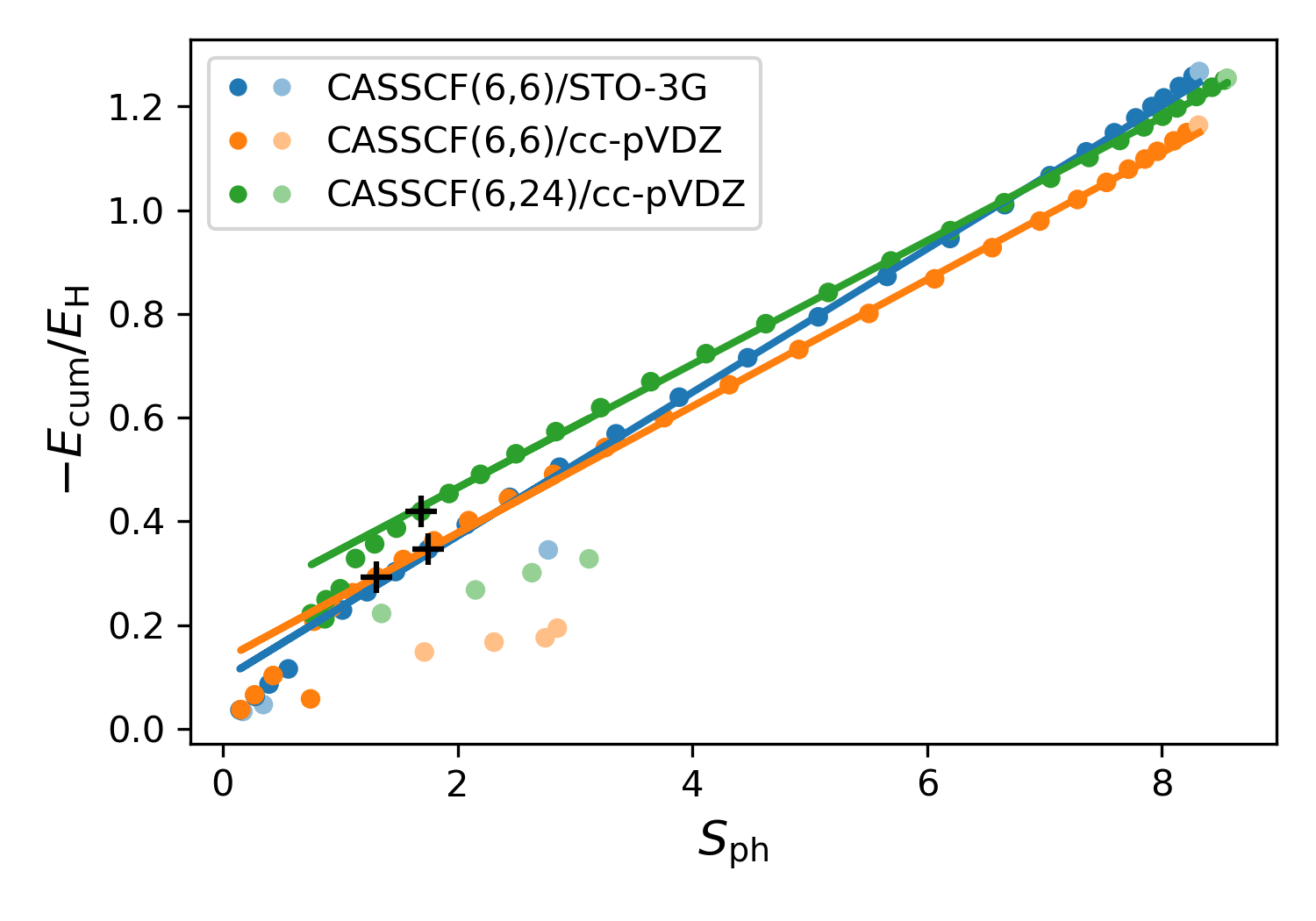}
\caption{Cumulant energy $\Ecum$ vs particle-hole symmetrized entropy $\Sph$ for the dissociation of dinitrogen.}
\label{fig:n2results}
\end{figure}

One of the most stable diatomics is N$_2$, which has a $^1\Sigma_g^+$ single-reference ground state at its equilibrium bond length and dissociates homolytically into two neutral nitrogen atoms upon bond stretching. The results for the dissociation are shown in Table~\ref{tab:n2results} and Fig.~\ref{fig:n2results}. As before, the equilibrium distance is marked with a full circle in the figure, and outlier points arising from changes in the electronic state at short bond lengths (for $R<0.73$~bohr in CASSCF(6,6) calculations and $R<0.98$~bohr in CASSCF(6,24) calculations), which were excluded from the fitting procedure, are shown in a lighter color.

\begin{table*}
\caption{Fit results from the dissociation curve and the $\Sph$--$\Ecum$ dependence for dinitrogen{, evaluated using the CASSCF$(n_\mathrm{e},n_\mathrm{orb})$ method}.}\label{tab:n2results}
\begin{tabular}{lcclllrrr}\toprule
Basis&${n_\mathrm{e}}$&${n_\mathrm{orb}}$&$D_\mathrm{e}$/$E_\mathrm{H}$&$r_\mathrm{e}$/bohr&$\tilde{\nu}$/cm$^{-1}$&$\kappa$/$E_\mathrm{H}$&$b$/$E_\mathrm{H}$&$\bar{\Delta}$/m$E_\mathrm{H}$\\\midrule
STO-3G&6&6&0.223&2.25&2085&0.139&0.0953&19.5\\
STO-3G&8&7&0.232&2.25&2097&0.136&0.110&16.8\\
STO-3G&10&8&0.239&2.25&2102&0.136&0.117&18.1\\
cc-pVDZ&6&6&0.314&2.10&2372&0.122&0.133&24.2\\
cc-pVDZ&6&12&0.323&2.09&2419&0.120&0.178&15.6\\
cc-pVDZ&6&18&0.341&2.10&2383&0.115&0.223&18.5\\
cc-pVDZ&6&24&0.330&2.11&2368&0.119&0.226&18.7\\
cc-pVDZ&8&8&0.332&2.10&2435&0.120&0.155&20.1\\
cc-pVDZ&8&14&0.346&2.10&2390&0.117&0.231&14.9\\
cc-pVDZ&8&20&0.367&2.10&2407&0.115&0.311&19.5\\\midrule
Exp.&&&0.364~\cite{lofthus77}&2.07~\cite{huber79}&2359~\cite{irikura07}\\
\bottomrule
\end{tabular}
\end{table*}

In terms of symmetry-adapted natural orbitals, heterolytic dissociation implies that each of the six highest-lying valence orbitals becomes half-occupied in the dissociation limit, such that the entropy reaches three times the value of the dissociated hydrogen molecule, $12\ln 2 \approx 8.32$. An equivalent formulation is that the electron pair associated with each bond distributes equally over the corresponding bonding and antibonding orbitals, which become degenerate in the dissociation limit. This implies that the orbital space can be partitioned into pairs whose summed occupation numbers remain constant throughout the dissociation. As the examples in this section illustrate, such “minimal orbital pairing” behavior, previously discussed in the framework of the functional theory of natural occupation numbers~\cite{piris2018electron}, appears to be a necessary condition for the validity of the CC. Similar approximations underlie many traditional approaches based on valence bond theory as well as modern methods such as localised active space approaches~\cite{hermes20}.

The data points show an approximately linear relation between $\Sph$ and $\Ecum$, although a very shallow turning point appears around $R = 3.2$~bohr in all curves, possibly related to a change in dynamic correlation. As before, $\kappa$ changes when moving from the minimal STO-3G to the cc-pVDZ basis set, as already reported in the Supplementary Material of Ref.~\cite{WangBaerends22-PRL}, while $b$ increases with the inclusion of virtual orbitals. Consistent with six electrons being involved in the bond breaking, correlating more than six electrons does not significantly affect $\kappa$ and only slightly modifies $b$. Overall, the values of $\kappa$ and $b$ are larger than in dihydrogen, but there is no simple quantitative relation between the two systems, even though the entropy behaves as if three independent bond dissociations were taking place. This confirms that these parameters vary across different covalent bonds.

As other homodiatomics, such as B$_2$, C$_2$, and O$_2$ discussed in the following, have a multireference character even at equilibrium distance. There are also low-lying excited states of singlet and triplet spin: they allow us to investigate the validity of the CC for systems that are multiconfigurational throughout their entire dissociation curve.

The ground state of C$_2$ is $^1\Sigma_g^+$; O$_2$ and B$_2$ have a $^3\Sigma_g^-$ ground state. To understand the impact of spin state and symmetry on the validity of the CC, we decided to investigate the lowest-energy state of both these symmetries for each of the molecules along their dissociation, even though it should be noted that $^1\Sigma_g^+$ is not the lowest-energy singlet state of dioxygen~\cite{huber79} and $^3\Sigma_g^-$ is not the lowest-energy triplet state of dicarbon~\cite{gurvich89}. 

\begin{figure}
\includegraphics[width=\linewidth]{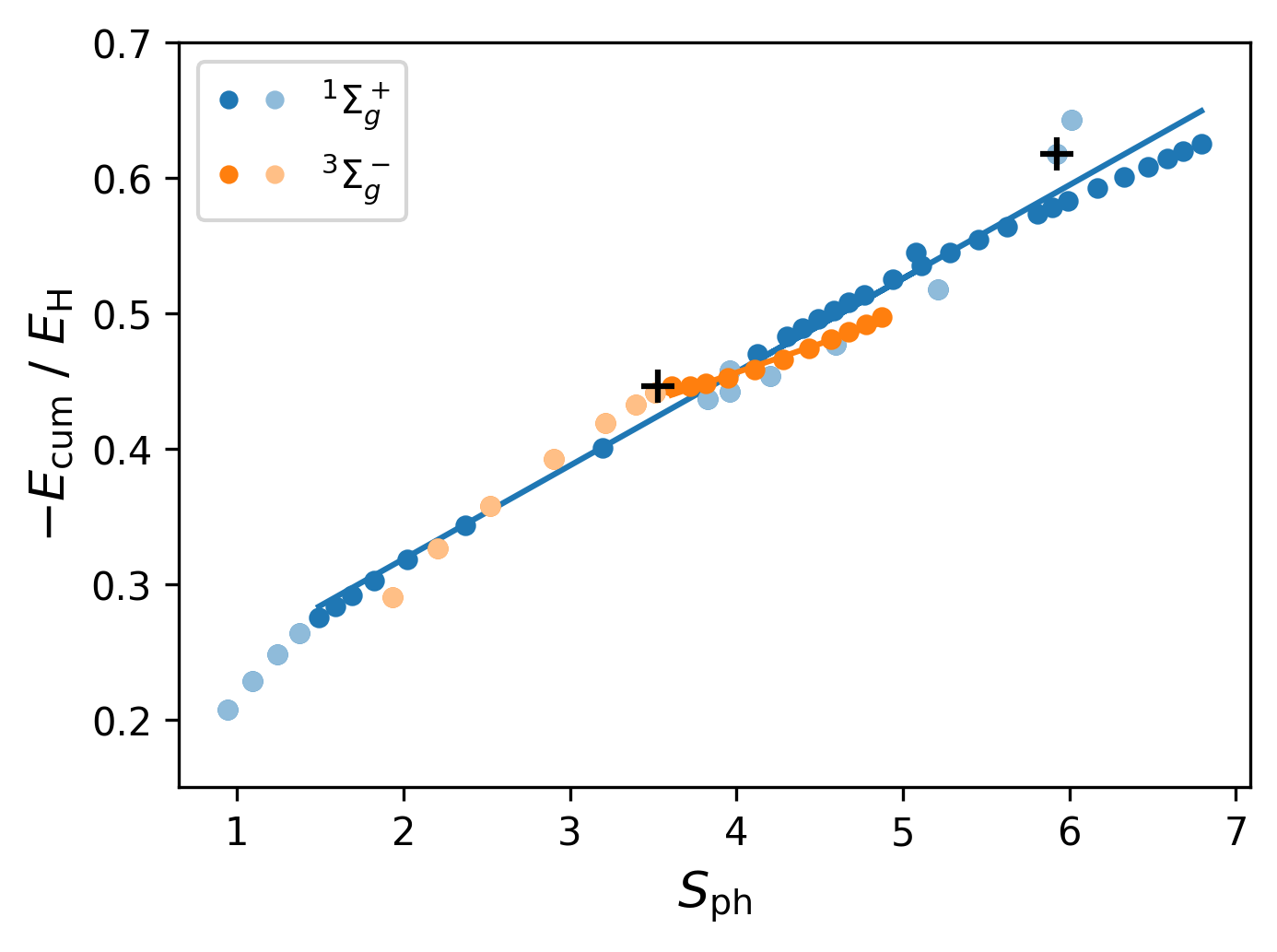}
\caption{Cumulant energy $\Ecum$ vs particle-hole symmetrized entropy $\Sph$ for the dissociation of diboron in the $^1\Sigma_g^+$ and $^3\Sigma_g^-$ states{, computed with CASSCF(6,26)/cc-pVDZ for both spin states.}}\label{fig:b2}
\end{figure}

\begin{figure}
\includegraphics[width=\linewidth]{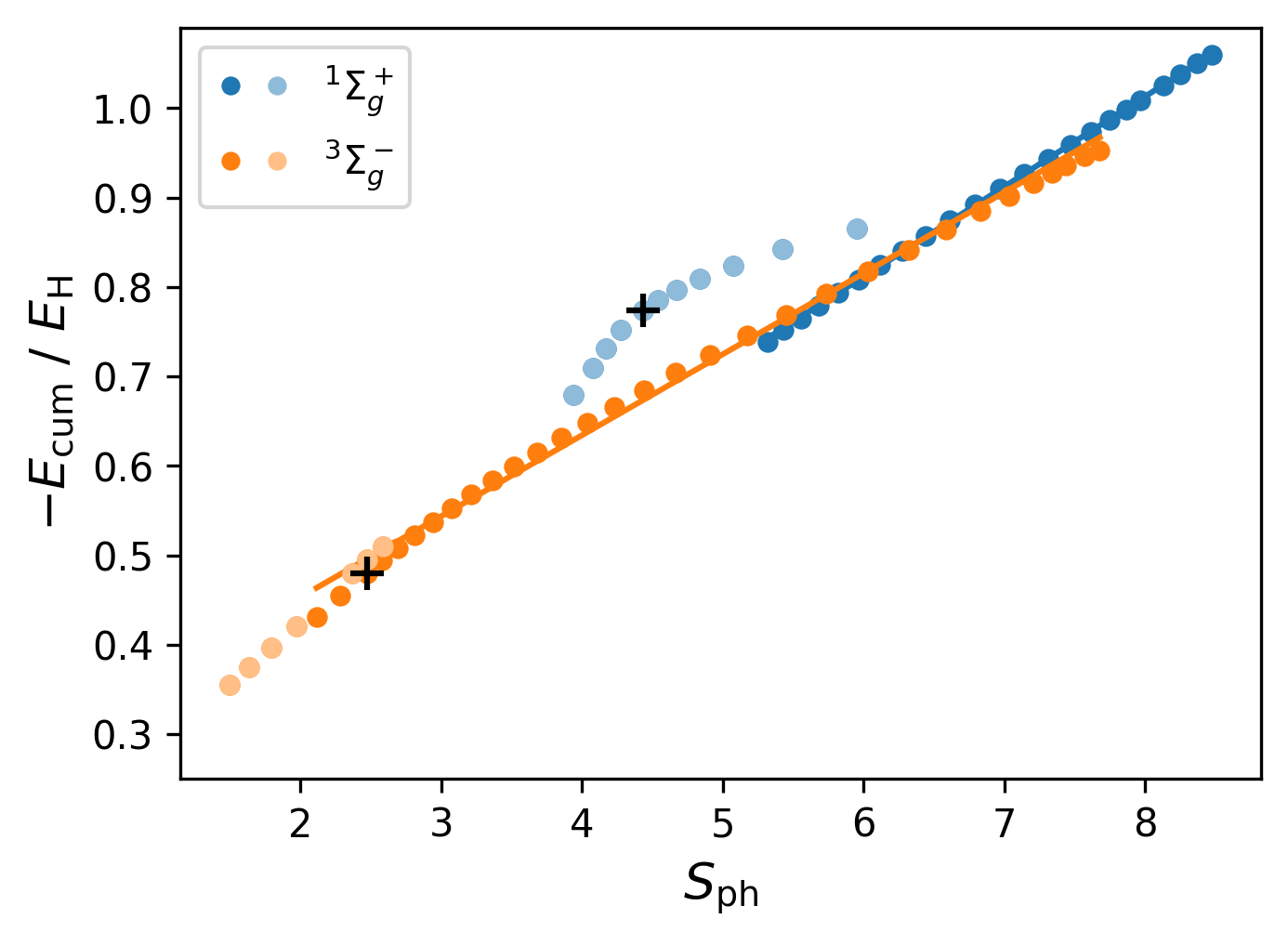}
\caption{Cumulant energy $\Ecum$ vs particle-hole symmetrized entropy $\Sph$ for the dissociation of dicarbon in the $^1\Sigma_g^+$ and $^3\Sigma_g^-$ states{, computed with CASSCF(8,26)/cc-pVDZ for both spin states.}}\label{fig:c2}
\end{figure}

\begin{figure}
\includegraphics[width=\linewidth]{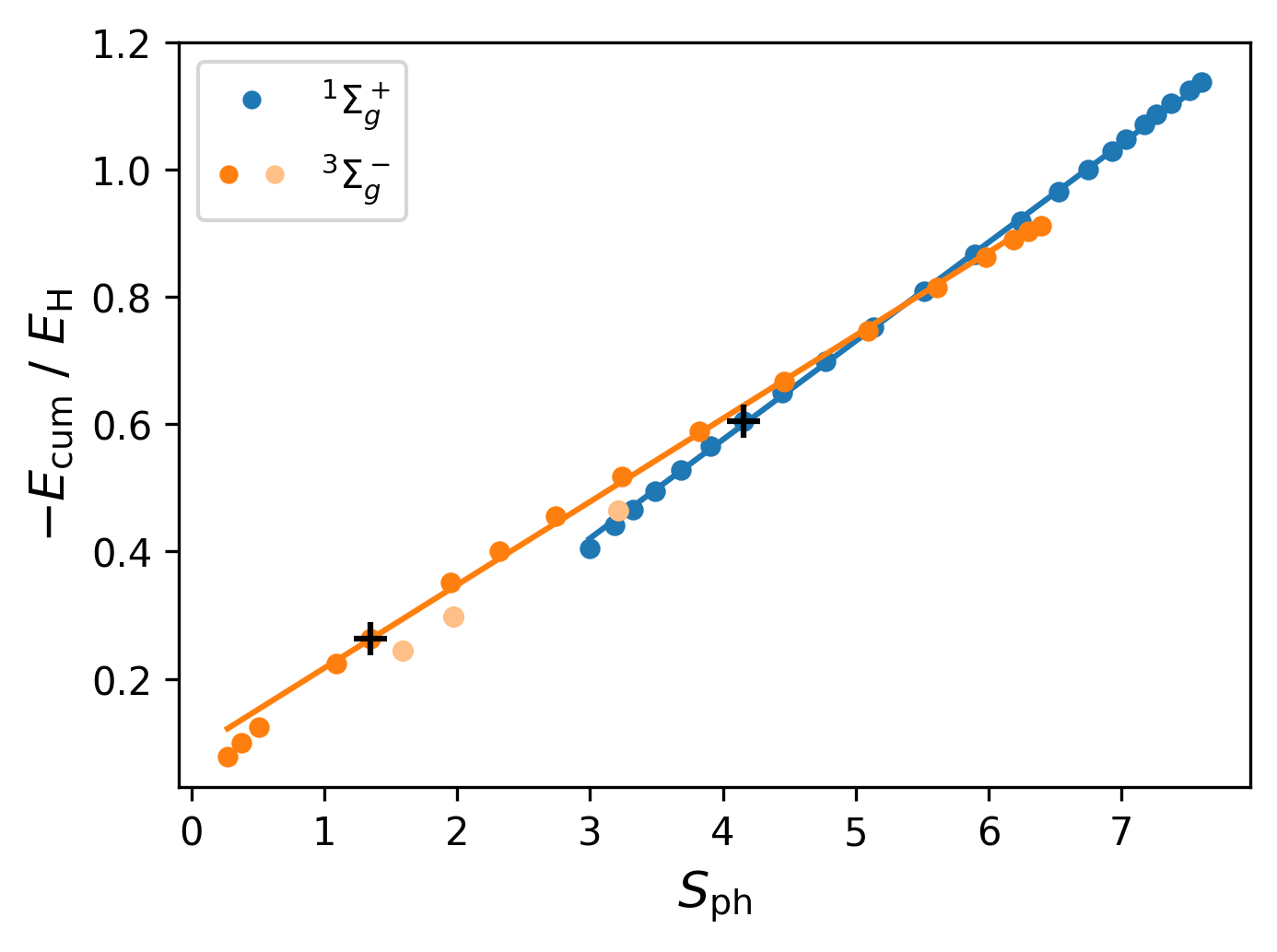}
\caption{Cumulant energy $\Ecum$ vs particle-hole symmetrized entropy $\Sph$ for the dissociation of dioxygen in the $^1\Sigma_g^+$ and $^3\Sigma_g^-$ states{, computed with CASSCF(16,10)/STO-6G for both spins states.}}\label{fig:o2}
\end{figure}

The results for a selection of basis sets and active spaces are shown in Table~\ref{tab:diatomics} and Figs.~\ref{fig:b2}-\ref{fig:o2}, which use the same color scheme as the previous plots: outliers that were not considered during the fitting procedure are shown in a lighter color. We attribute such outliers to jumps in the electronic configuration or to a violation of the minimal orbital pairing behavior, as detailed below. The ranges of $R$ that were used in the fitting procedure as well as the dependence of $\Sph$ on $R$ are available in the supplementary material. It should be pointed out that for the singlet states of B$_2$ and C$_2$, the electronic state at equilibrium only persists for a small range of $R$ and furthermore does not follow the CC; these data points were not included in the fitting procedure.

As in the previous examples, an accurate estimation of $\kappa$ is only possible with cc-pVDZ or larger basis sets, and accurately estimating $b$ requires an active space beyond only the valence orbitals. For this reason, the calculations for B$_2$ and C$_2$ were carried out with all valence and virtual orbitals active (frozen core-full configuration interaction, fc-FCI) in the cc-pVDZ basis set. Due to the high computational requirements, the calculations for O$_2$ were instead carried out as FCI calculations in the STO-3G basis set, which might introduce an inaccuracy to $\kappa$ but should make it possible to qualitatively assess the CC and compare it to previous results~\cite{Irimia2023-open-shell}.

\begin{table*}
\caption{Fit results from the dissociation curve and the $\Sph$--$\Ecum$ dependence for several diatomic molecules{, evaluated using the CASSCF$(n_\mathrm{e},n_\mathrm{orb})$ method}.}\label{tab:diatomics}
\begin{tabular}{lccccllll@{\hspace{1ex}}rl}\toprule
Molecule & State & Basis & $n_\mathrm{e}$ & $n_\mathrm{orb}$ & $D_\mathrm{e}$/$E_\mathrm{H}$ & $r_\mathrm{e}$/bohr & $\tilde{\nu}$/cm$^{-1}$ & $\kappa$/$E_\mathrm{H}$ & $b$/$E_\mathrm{H}$ & {$\bar{\Delta}$/m$E_\mathrm{H}$}\\\midrule
& $^1\Sigma_g^+$ & cc-pVDZ & 6 & 26 & 0.071 & 3.095 & 956 & 0.0673 & 0.184   & 7.05 \\
B$_2$ & $^3\Sigma_g^-$ & cc-pVDZ & 6 & 26 & 0.094 & 3.074 & 1031 & 0.0423 & 0.286   & 2.77 \\
& $^3\Sigma_g^-$ & Exp. & & & 0.112(22)~\cite{huber79,irikura07} & 3.00~\cite{gurvich89} & 1060~\cite{irikura07} &\\\midrule
\multirow{8}{*}{C$_2$} & $^1\Sigma_g^+$ & STO-3G&8&8&0.253&2.36&1472&0.120&--0.00664&4.37\\
& $^1\Sigma_g^+$ & cc-pVDZ&4&24&0.159&2.66&1347&0.0894&0.0699&5.39\\
& $^1\Sigma_g^+$ & cc-pVDZ&6&24&0.233&2.36&1879&0.0919&0.135&0.347\\
& $^1\Sigma_g^+$ & cc-pVDZ&8&8&0.221&2.37&1887&0.0934&0.117&1.83\\
& $^1\Sigma_g^+$ & cc-pVDZ & 8 & 26 & 0.206 & 2.388 & 1788 & 0.100  & 0.206   & 1.21 \\
& $^1\Sigma_g^-$ & Exp. & & & 0.231~\cite{uhndal91} & 2.35~\cite{huber79} & 1855~\cite{huber79} \\
& $^3\Sigma_g^-$ & cc-pVDZ & 8 & 26 & 0.178 & 2.642 & 1413& 0.0905 & 0.252   & 13.6 \\
& $^3\Sigma_g^-$ & Exp. & & & 0.189~\cite{mueller00} & 2.59~\cite{davis88} & 1470~\cite{davis88} \\\midrule
\multirow{6}{*}{O$_2$} & $^3\Sigma_g^-$ & STO-6G  & 16& 10 & 0.143 & 2.433 & 1671 & 0.130  & 0.0862  & 11.9 \\
& $^3\Sigma_g^-$ & Exp. & & & 0.192~\cite{huber79} & 2.282~\cite{huber79} & 1580~\cite{huber79} \\
& $^1\Sigma_g^+$ & STO-6G  & 16& 10 & 0.108 & 2.471 & 1440 & 0.154  & --0.0441 & 4.86 \\
& $^1\Sigma_g^+$ & cc-pVDZ&8&6&0.101&2.32&1454&0.140&--0.00972&8.94\\
& $^1\Sigma_g^+$ & cc-pVDZ&8&20&0.127&2.27&1631&0.121&0.249&5.04\\
& $^1\Sigma_g^+$ & Exp. & & & 0.132~\cite{huber79} & 2.297~\cite{huber79} & 1433~\cite{huber79}\\
\bottomrule
\end{tabular}
\end{table*}

The comparison of the three molecules reveals similarities in their $\Sph$--$\Ecum$ relations. In all cases, the entropy and the cumulant energy increase monotonously over the course of the dissociation as long as the system stays within one electronic state and each curve possesses at least one linear part. In these, fits always result in a smaller $\kappa$ and a larger $b$ in the triplet state.

Notably, the occupation number pattern in these molecules and thus the entropy they reach at dissociation cannot be predicted based on the redistribution of electrons within pairs of occupied and unoccupied orbitals according to ``minimal orbital pairing''. This is because at low bond distances, $\upsigma$ and $\uppi$ orbitals differ in energy, but they change into degenerate p orbitals during dissociation which are equally occupied. This is demonstrated in Table~\ref{tab:minimalpairing} where we show the approximate vectors of natural occupation numbers of the highest-energy valence orbital and the corresponding particle-hole symmetrized entropies for different diatomic species in their singlet state before and after dissociation. In the second-row element diatomics, these orbitals are listed in the order $3\upsigma_g = \upsigma_{\mathrm{2p}_z}$,$1\uppi_u^x = \uppi_{\mathrm{2p}_x}$,$1\uppi_u^y = \uppi_{\mathrm{2p}_y}$,$1\uppi_g^x = \uppi^*_{\mathrm{2p}_x}$,$1\uppi_g^y = \uppi^*_{\mathrm{2p}_y}$,$3\upsigma_u = \upsigma^*_{\mathrm{2p}_z}$. It should be noted that while the ground state of B$_2$ at equilibrium distance (which is not considered in the fitting procedure) has spin-summed occupation numbers, NON(ss), of (0,1,1,0,0,0), it is soon crossed by the (2,0,0,0,0,0) state that is the ground state over most of the dissociation.

\begin{table*}
    \centering
    \caption{Patterns of spin-summed occupation numbers of the highest-energy spatial valence orbitals of different diatomic molecules a) before dissociation, b) in the dissociation limit according to ``minimal orbital pairing'', and c) in the physical ground state in the dissociation limit.}
    \begin{tabular}{lcc|cc|cc}\toprule
        &\multicolumn{2}{c}{a)}&\multicolumn{2}{c}{b)}&\multicolumn{2}{c}{c)}\\
        &NON(ss)&$\Sph$&NON(ss)&$\Sph$&NON(ss)&$\Sph$\\\midrule
        H$_2$ (singlet)&2,0&0&1,1&$4\ln 2 \approx 2.77$&1,1&$4\ln 2 \approx 2.77$\\
        N$_2$ (singlet)&2,2,2,0,0,0&0&1,1,1,1,1,1&$12\ln 2 \approx 8.32$&1,1,1,1,1,1&$12\ln 2 \approx 8.32$\\[1.5ex]
        B$_2$ (singlet)&2,0,0,0,0,0&$0$&1,0,0,0,0,1&$4\ln 2\approx 2.77$&$\frac{1}{3},\frac{1}{3},\frac{1}{3},\frac{1}{3},\frac{1}{3},\frac{1}{3}$&$12\ln 2+12\ln 3-10\ln 5 \approx 5.406$\\[1.5ex]
        B$_2$ (triplet)&0,1,1,0,0,0&$4\ln 2 \approx 2.77$&$0,\frac{1}{2},\frac{1}{2},\frac{1}{2},\frac{1}{2},0$&$16\ln 2-6\ln 3\approx 4.50$&$\frac{1}{3},\frac{1}{3},\frac{1}{3},\frac{1}{3},\frac{1}{3},\frac{1}{3}$&$12\ln 2+12\ln 3-10\ln 5 \approx 5.406$\\[1.5ex]
        O$_2$&2,2,2,1,1,0&$4\ln 2 \approx 2.77$&$1,\frac{3}{2},\frac{3}{2},\frac{3}{2},\frac{3}{2},1$&$20\ln 2-6\ln 3 \approx 7.27$&$\frac{4}{3},\frac{4}{3},\frac{4}{3},\frac{4}{3},\frac{4}{3},\frac{4}{3}$&$12\ln 3-8\ln 2 \approx 7.64$\\[1.5ex]
        C$_2$&2,1,1,0,0,0&$4\ln 2 \approx  2.77$&$1,\frac{1}{2},\frac{1}{2},\frac{1}{2},\frac{1}{2},1$&$20\ln 2-6\ln 3 \approx 7.27$&$\frac{2}{3},\frac{2}{3},\frac{2}{3},\frac{2}{3},\frac{2}{3},\frac{2}{3}$&$12\ln 3-8\ln 2 \approx 7.64$\\\bottomrule
    \end{tabular}
    \label{tab:minimalpairing}
\end{table*}

The deviation from minimal orbital pairing is a possible reason for the different extent of agreement with the CC in each molecule. Singlet B$_2$ shows the most deviations from linearity of the three molecules: there are two significant bends of the curve at entropies of 1.5 ($R = 1.2$~bohr) and 4.5 ($R = 5.9$~bohr). This could be related to a strong reorganization of electrons among different orbital pairs: the entropy increases to almost twice the value that could result within a minimal pairing approach. In the triplet state of the same molecule, the 3$\sigma_g$ orbital cannot be doubly occupied. For this reason, the expected entropy according to minimal pairing is higher, making the difference to the physical entropy in the dissociation limit smaller. This is in line with the $\Sph$--$\Ecum$ curve in the triplet state, which resembles two linear segments with only one bend near the equilibrium structure.

In contrast, the computed relations in C$_2$ and O$_2$ are almost perfectly linear. This holds for both triplet states, as confirmed for O$_2$ in a previous study~\cite{Irimia2023-open-shell}, and singlet states. Even though they do not follow the minimal pairing approach completely, the extent of electron reorganization is smaller, which explains the linear behavior over a large part of the dissociation curves. The only deviations from linearity come about due to jumps between energy surfaces.

An example of a heterodiatomic molecule that is isoelectronic to dinitrogen is carbon monoxide. Its triple bond represents the strongest chemical bond of all molecules, and while the oxygen atom formally carries a positive charge, the dissociation of the bond does not proceed homolytically. Instead, one of the electron pairs localizes on oxygen to produce the two neutral atoms as fragments. Thus, it represents an interesting intermediate example between most of the examples for homolytic dissociation in this paper, and helium hydride which dissociates under electron localization and was previously found to not fulfill the CC.

\begin{figure}
\includegraphics[width=\linewidth]{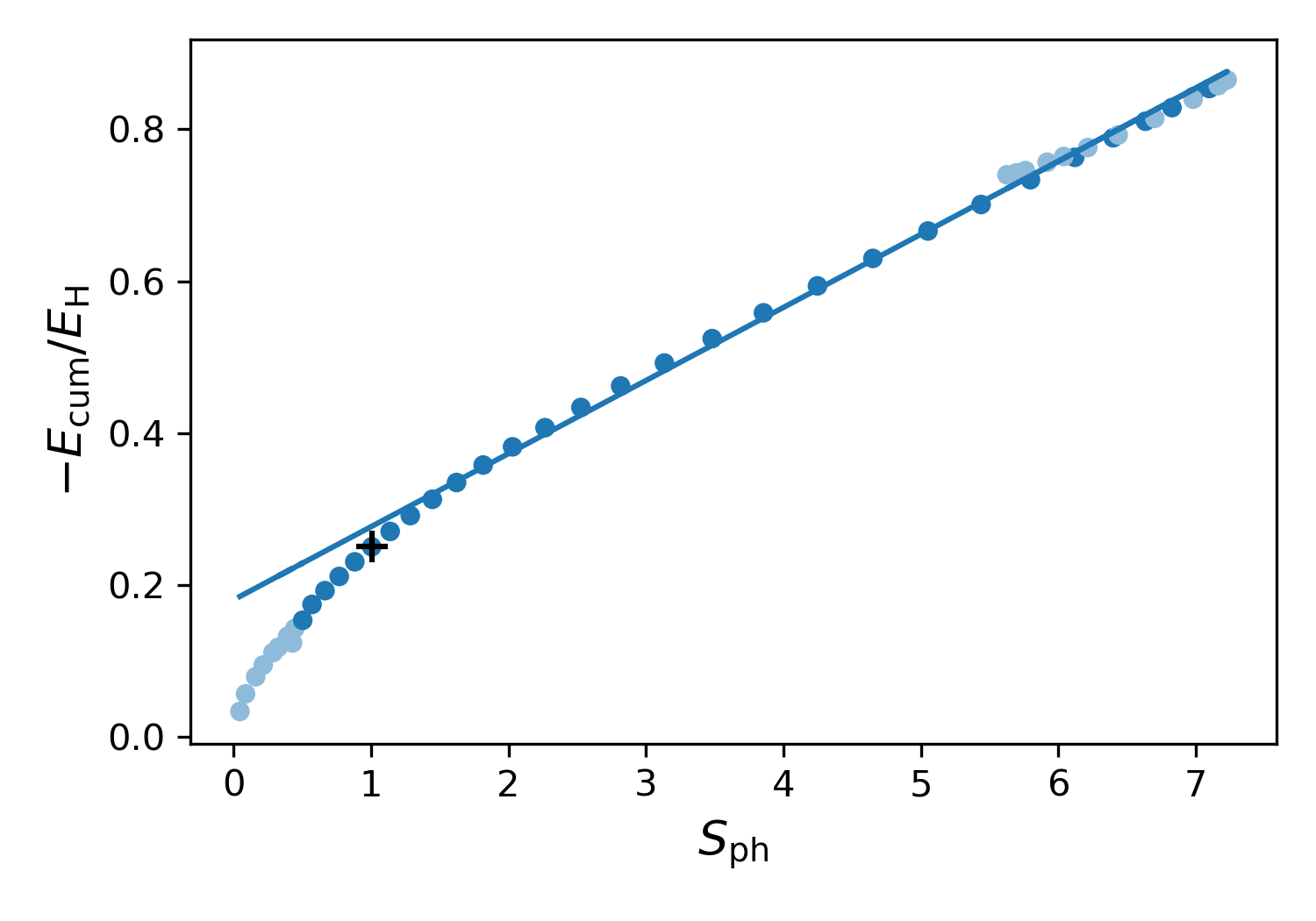}
\caption{Cumulant energy $\Ecum$ vs particle-hole symmetrized entropy $\Sph$ for the dissociation of carbon monoxide{, computed with CASSCF(6,6)/cc-pVDZ}.}\label{fig:coresults}
\end{figure}

The $\Sph$--$\Ecum$ relation obtained with CASSCF(6,6)/cc-pVDZ is shown in Fig.~\ref{fig:coresults}. The linear fit to the data with $0.60$~bohr $\leq R < 5.49$~bohr shown in the plot results in $\kappa=0.0962 E_\mathrm{H}$ and $b = 0.181 E_\mathrm{H}$. Data for other basis sets are available in the supplementary material.

In line with the six electrons involved in the bonding, at least six electrons need to be explicitly correlated in order to obtain a good approximation for $\kappa$. In such calculations, the curves obtain noticeable deviations from linearity. CASSCF(8,6) calculations seem to deviate strongly from the pattern, while those with ten active electrons yield more similar results to those with six active electrons. Following the general trend, cc-pVDZ calculations give lower values for $\kappa$ compared to STO-3G calculations with the same number of active electrons, but yield much larger values for $b$ when including dynamic correlation. In general, the values for $\kappa$ are smaller, but in the same order of magnitude compared to dinitrogen.

The plot of the cumulant energy vs the entropy shows a turning point that can be related to the electronic structure of carbon monoxide in the dissociation limit. While three covalent bonds are stretched and the electrons distribute over bonding and antibonding orbitals, the entropy approaches the theoretical limit of $12\ln 2 \approx 8.32$ in the CASSCF(6,6) calculation. At $R = 5.3$~bohr, the entropy reaches a maximum and decreases again as one of the electron pairs becomes localized on the oxygen atom. At this point, only four electrons are pairwise entangled and four orbitals approach half-occupation in the limit, such that the entropy approaches $8\ln 2 \approx 5.55$. While this change leads to a deviation from the linear behavior, the $R$ values at which the maximum values of $\Sph$ and $\Ecum$ are reached coincide and the slope changes only gradually for larger $R$.

At this point, it can be concluded that the CC holds for a class of dissociating diatomic molecules, both symmetric and asymmetric. Exceptions have been found for helium hydride, which has a mostly coordinative bond that dissociates heterolytically, and for those second-row diatomics that change electronic state during dissociation or that do not fulfill a minimal pairing conjecture. In these cases, the CC typically only holds along some part of the dissociation trajectory.

\subsection{Polyatomic molecules}

The electronic structure of molecules with more than two atoms is generally more complex than that of diatomics. In particular, these can have different types of bonds with different strengths, that interact with each other and can dissociate in different ways. Furthermore, there are nondissociative degrees of freedom such as bending and torsion that can modify the bonding situation in the molecule, which also comprise most chemical elementary reactions.

In this section, we investigate the validity of the CC for different degrees of freedom in four different molecules that are representative of many common inorganic and organic molecules but small enough to conduct high-level calculations. In this setting, an exact behavior according to the conjecture would imply that information on the complete potential energy surface of a molecule can be derived by analyzing an infinitesimal deformation along a single degree of freedom.

The validity of the CC in the water molecule has previously been confirmed~\cite{WangKnowles21-PRA} for bond angles of 80$^\circ$--120$^\circ$ and bond lengths of 1.5--5~bohr on the CASSCF(6,14)/cc-pVDZ level of theory. In this work, we investigate three deformations of water and the isoelectronic hydrogen sulfide separately over large ranges: the dissociation of a single H atom, the symmetric dissociation of both H atoms, and the variation of the bond angle. These calculations were carried out at the CASSCF(8,23)/cc-pVDZ (=fc-FCI) level of theory.

\begin{figure}
\includegraphics[width=\linewidth]{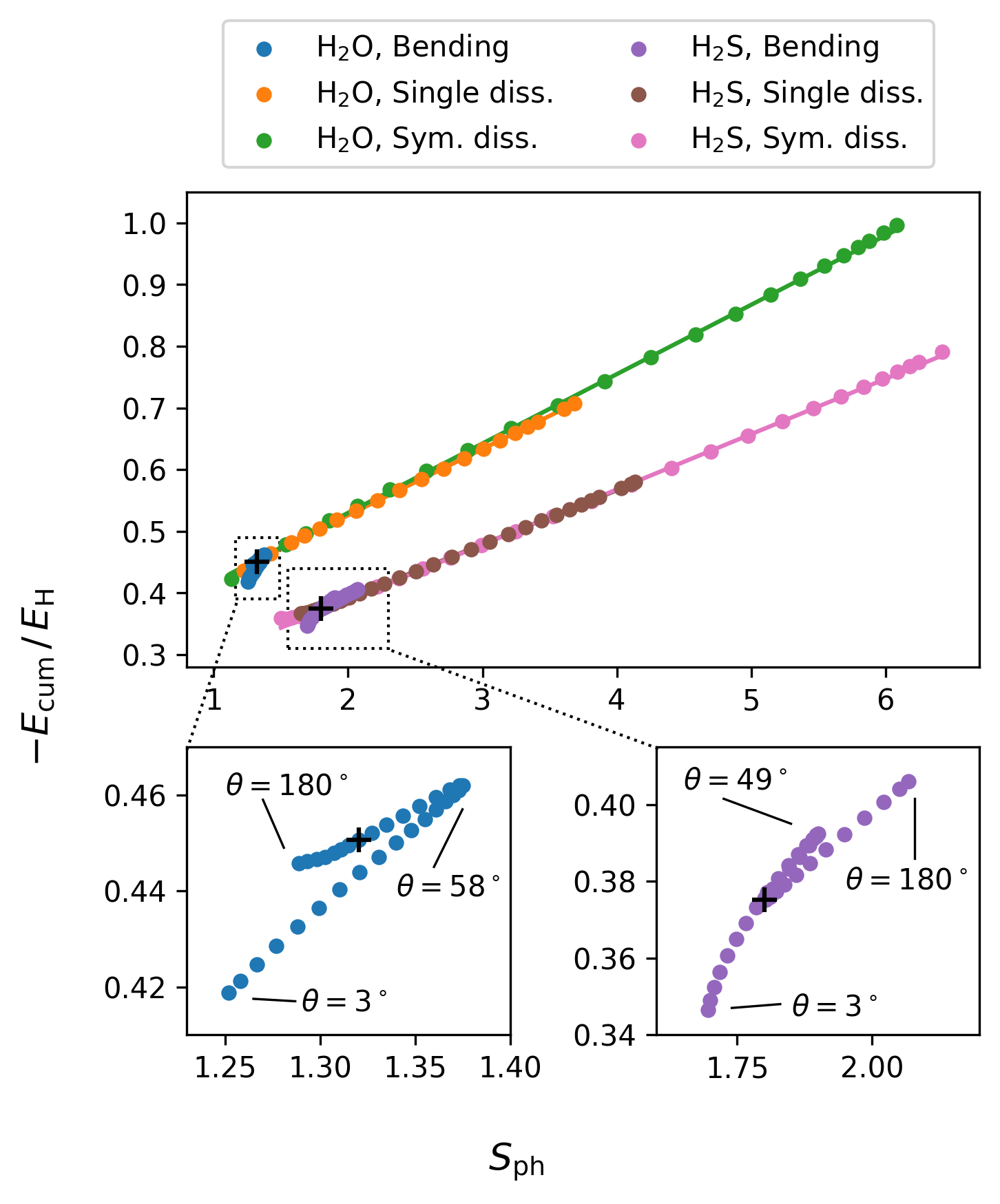}
\caption{H$_2$O and H$_2$S. Cumulant energy $\Ecum$ vs particle-hole symmetrized entropy $\Sph$ for the dissociation of one H atom, the symmetric dissociation of both H atoms, and the bending motion in water and hydrogen sulfide{, computed with CASSCF(8,23)/cc-pVDZ for all processes in both molecules}.}\label{fig:h2oh2s}
\end{figure}

The results for the $\Sph$--$\Ecum$ relation for water and hydrogen sulfide are shown in Fig.~\ref{fig:h2oh2s}. The data obtained from fitting these curves as well as the dissociation curve are available from Table~\ref{tab:large_mol_results}.

Both single and symmetric dissociation curves for both molecules show linearity with weighted RMSE values of only 1 to 4~m$E_\mathrm{H}$. This is linearity comparable to our results for hydrogen and better than those for diatomics. The slight deviations from linearity appear at very small bond lengths, similar to N$_2$ or H$_2$. Furthermore, the curves lie almost on top of each other, i.e., almost identical $\kappa$ and $b$ result for the two processes. This shows how despite a two-dimensional parametrization of the molecular geometry, the CC which reduces the system to the one-dimensional $\Sph$, still applies. In the dissociation limit, the curves approach the theoretical entropy limits for the dissociation of one ($4\ln 2\approx 2.77$) and two ($8\ln 2\approx 5.55$) bonds, implying that each bond's electrons reorganize within an orbital pair during dissociation. 

Upon bending, cumulant energy and entropy change over a much smaller range: Most of the energy change due to bending seems to be a purely mean-field effect as the correlation energy and the natural occupation numbers do not change much. However, there is still a structure in the $\Sph$--$\Ecum$ curves: In water, the cumulant energy decreases with the entropy upon opening the bond toward a linear geometry, with a slope similar to that of the bond dissociation curves. When reducing the angle between the O--H-bonds, the cumulant energy linearly increases with the entropy until a turning point is reached around an angle of 58$^\circ$, after which the cumulant energy linearly decreases again with a different slope. A potential explanation for this is that this deformation pushes the hydrogen atoms together, creating a bonding interaction between them, and the electrons become increasingly localized in this bond instead of in the two hydrogen--oxygen bonds. The non-linearity is once again caused by this beyond-minimal pairing reorganization of electron density, but after that point the system fulfills the CC (just as confirmed for isolated H$_2$ in Sec.~\ref{sec:secondrow}), yet with different slopes depending on the presence of an atom in the environment.

A similar non-monotonous curve results in hydrogen sulfide, however with two extrema. Compressing the bond angle first increases entropy and cumulant energy in a linear fashion until a turning point at 49$^\circ$. After that, the quantities decrease but curve slightly when the hydrogen atoms approach each other. When increasing the angle, there is first a local minimum in entropy and cumulant energy at 104$^\circ$, followed by a linear increase of cumulant energy with entropy, where $\Sph$ increases significantly by 0.25.

\begin{figure}
\includegraphics[width=\linewidth]{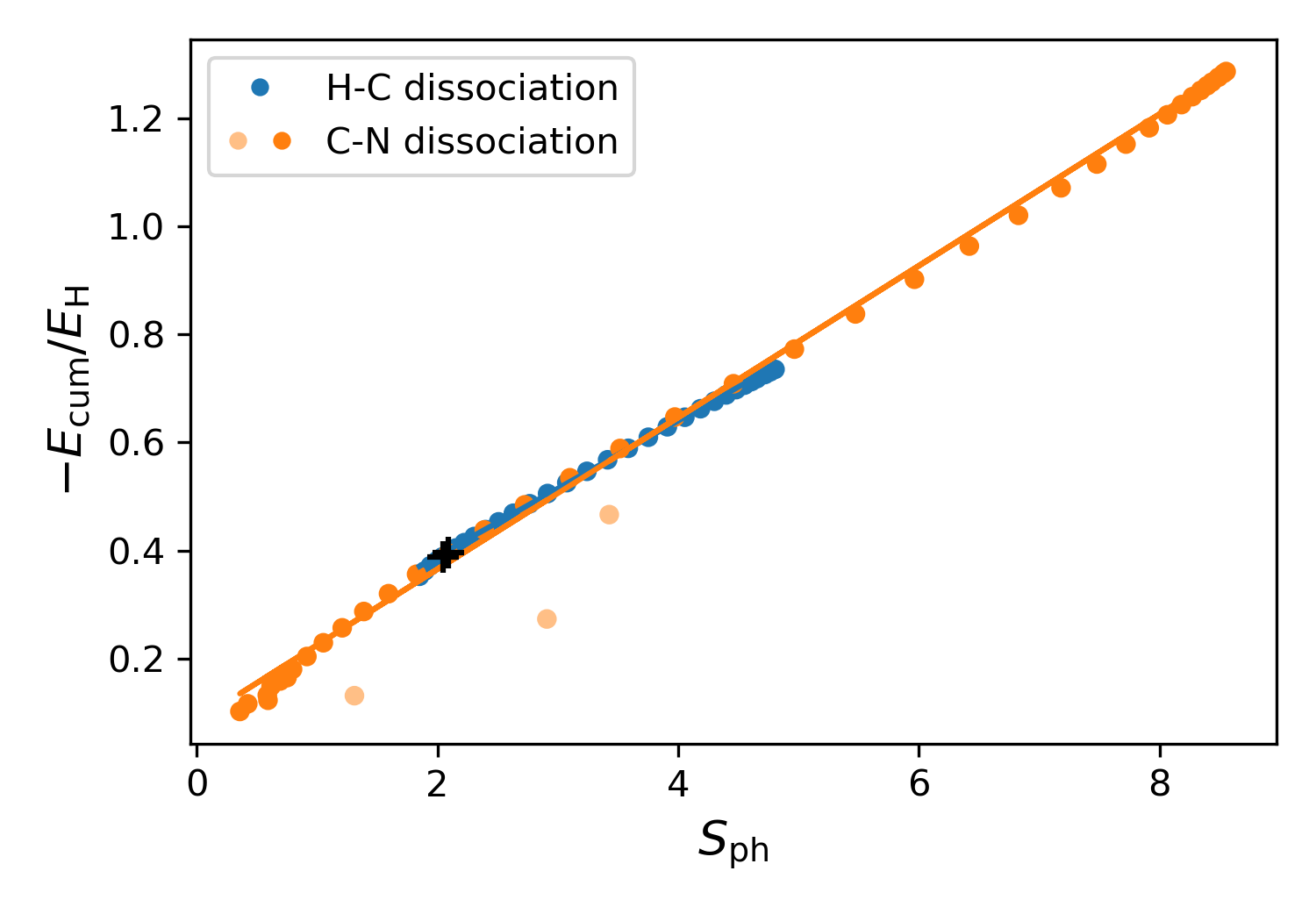}
\caption{Cumulant energy $\Ecum$ vs particle-hole symmetrized entropy $\Sph$ for the H--C vs. the C--N-bond dissociation in hydrogen cyanide{, computed with CASSCF(10,9)/STO-3G}.}\label{fig:hcn}
\end{figure}

The hydrogen cyanide molecule is a simple example of a molecule with two different types of bonds: a strong triple bond between carbon and nitrogen and a single bond between hydrogen and carbon. Maintaining the linear arrangement of the three atoms, we compare the dissociation of the two bonds in two series of calculations. Since fc-FCI calculations in cc-pVDZ were not feasible, we instead performed a series of calculations with different numbers of active electrons and all virtual orbitals active in STO-3G. Fig.~\ref{fig:hcn} shows the results of fc-FCI calculations in the STO-3G basis set. The fits in this curve result in $\kappa =$ 0.125~$E_\mathrm{H}$ and $b =$ 0.139~$E_\mathrm{H}$ for the dissociation of the hydrogen--carbon bond, and $\kappa = $0.135~$E_\mathrm{H}$ and $b =$ 0.104~$E_\mathrm{H}$ for the carbon--nitrogen bond. The results of the fits with smaller active spaces are available in the supplementary material.

These calculations confirm a linear behavior for both dissociation processes, with a weak S-shape for the dissociation of the C--N-bond and a few outliers at small bond distances. The $\kappa$ of these curves can already be approximately reproduced with two active electrons for the H--C-dissociation and six active electrons for the C--N-dissociation, in line with the number of electrons involved in these bonds. Just as discussed in Sec.~\ref{sec:two_elec_sys}, further increasing the number of active electrons only increases $b$. Notably, the parameters of the very simple CC expression are the same for these two different and asymmetric degrees of freedom of the molecular geometry.

To further substantiate this insight on the universality of $\kappa$ and $b$ for this molecule, we repeated the calculations with the cc-pVDZ basis set with limited numbers of active electrons and virtual orbitals. The results for $\kappa$ and $b$ are available in the supplementary material. They confirmed the sensitivity of $\kappa$ to the basis set choice and showed that they indeed depend on the dissociating bond in the larger basis set: STO-3G overestimates $\kappa$ by up to 20\% compared to cc-pVDZ for the C--N-dissociation and by 43\% for the H--C-dissociation, while the conformity to the linear fit is not affected.

Finally, we investigated the ethylene molecule, a simple example of an unsaturated organic molecule. In carbon--carbon double bonds, bond breaking can not only be elicited by dissociation, but also by torsion of the double bond. By changing the dihedral angle between the two CH$_2$ fragments, the overlap between the p orbitals that form the bond is reduced and the former $\uppi$ and $\uppi^*$ orbitals become very close in energy. In this way, static correlation is introduced without the need to approach a dissociation limit.

\begin{figure}
\includegraphics[width=\linewidth]{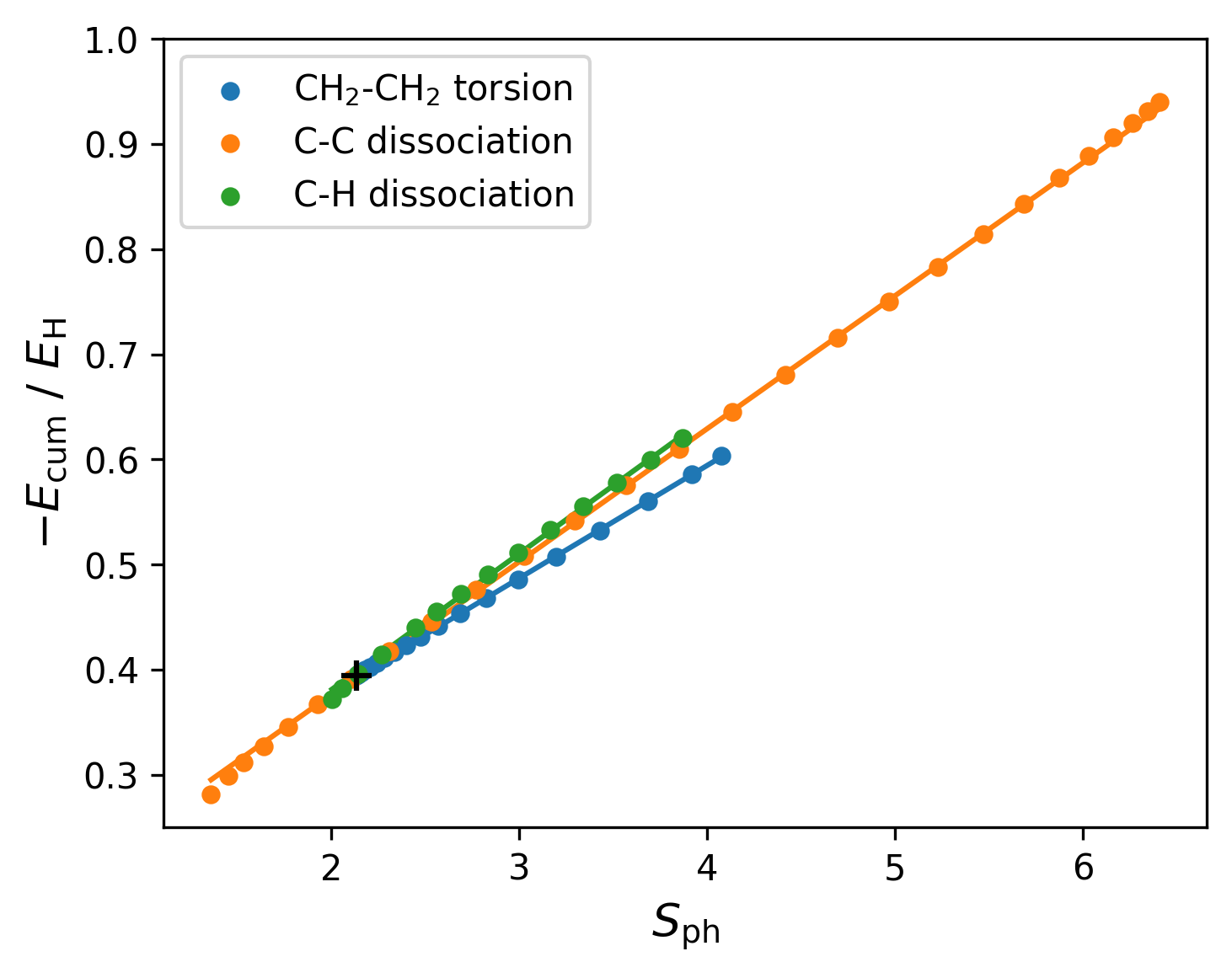}
\caption{Cumulant energy $\Ecum$ vs particle-hole symmetrized entropy $\Sph$ for three different geometric deformations of ethylene{, computed with CASSCF(12,12)/STO-6G.}} \label{fig:c2h4}
\end{figure}

In particular, we have studied three different deformations of an ethylene molecule: the dissociations of the H--C and the C--C bonds, and the torsion of the two CH$_2$ groups relative to each other.
The calculations were carried out at the FCI level in the STO-6G basis set. 
However, we expect based on the previous results that $\kappa$ can be reproduced with a 20\%-40\% tolerance and a linearity observed with STO-6G will also hold in the basis set limit. A graphical representation of the $\Sph$--$\Ecum$ dependencies is shown in Fig.~\ref{fig:c2h4} and the numerical data resulting from the fits are available from Table~\ref{tab:large_mol_results}. 

\begin{table*}
\caption{Fit results from the dissociation curves and the $\Sph$--$\Ecum$ dependencies for ethylene, water, and hydrogen sulfide{, evaluated using the CASSCF$(n_\mathrm{e},n_\mathrm{orb})$ method}.}\label{tab:large_mol_results}
\begin{tabular}{lcccclllrrr}\toprule
Molecule & Deformation & Basis & $n_\mathrm{e}$ & $n_\mathrm{orb}$ & $D_\mathrm{e}$/$E_\mathrm{H}$ & $r_\mathrm{e}$/bohr & $\tilde{\nu}$/cm$^{-1}$ & $\kappa$/$E_\mathrm{H}$ & $b$/$E_\mathrm{H}$ & $\bar{\Delta}$/m$E_\mathrm{H}$ \\\midrule
C$_2$H$_4$&C-C diss. &STO-6G&12&12&0.285&2.565&1571&0.126&0.123&3.06\\
& & Exp. & & & 0.286~\cite{chase82,furtenbacher06,lerberghe72}$^\mathrm{a}$ & 2.530~\cite{herzberg66} & 1623~\cite{lerberghe72}\\
          &C-H diss. &STO-6G&12&12&0.204&2.084&3385&0.129&0.122&2.48\\
& & Exp. & & & 0.186~\cite{atct,cccbdb,lerberghe72}$^\mathrm{a}$ & 2.052~\cite{herzberg66} & 3026~\cite{lerberghe72}$^\mathrm{b}$\\          
H$_2$O    & Single diss. &cc-pVDZ&8&23&0.183&1.827&3897&0.107&0.311&1.48\\
          & Symm. diss.  &cc-pVDZ&8&23&0.333&1.827&3972&0.112&0.305&2.95\\
          & & Exp. & & &0.1999\cite{chase82,irikura07,huber79,shimanouchi72}$^\mathrm{a}$&1.810~\cite{hoy79}&3832~\cite{shimanouchi72}$^\mathrm{b}$\\
H$_2$S     & Single diss.  &cc-pVDZ&8&23&0.140&2.524&2820&0.087&0.216&1.08\\
           & Symm. diss. &cc-pVDZ&8&23&0.262&2.519&2842&0.089&0.209&3.12\\
           & Exp. & & & 2.524~\cite{cook75} &0.1494~\cite{irikura07,chase82,shimanouchi72}$^\mathrm{a}$&&2722~\cite{shimanouchi72}$^\mathrm{b}$\\
           \midrule
Molecule & Deformation & Basis & $n_\mathrm{e}$ & $n_\mathrm{orb}$ & $\Delta E$/$E_\mathrm{H}$ & $\theta_\mathrm{e}$/$^\circ$ & $k_2$/$E_\mathrm{H}$Rad$^{-2}$ & $\kappa$/$E_\mathrm{H}$ & $b$/$E_\mathrm{H}$ & $\bar{\Delta}$/m$E_\mathrm{H}$ \\\midrule
C$_2$H$_4$ & Torsion  & STO-6G  & 12 & 12 & 0.123$^\mathrm{c}$ & 0     & 0.150 & 0.107 & 0.165 & 0.66\\
H$_2$O     & Bending & cc-pVDZ & 8  & 23 & 0.0625 & 102.1$^\mathrm{d}$ & 0.173 & -- & -- & -- \\
H$_2$S     & Bending & cc-pVDZ & 8  & 23 & 0.114 & 92.4$^\mathrm{d}$ & 0.178 & -- & -- & -- \\\bottomrule
\multicolumn{11}{l}{$^\mathrm{a}$Energy for the adiabatic dissociation of a single bond.}\\
\multicolumn{11}{l}{$^\mathrm{b}$Vibrational mode resembling the symmetric stretching.}\\
\multicolumn{11}{l}{$^\mathrm{c}$Exp.: 0.119~$E_\mathrm{H}$ from thermal rotation barrier~\cite{douglas55}, ZPE for C$_2$H$_4$~\cite{lerberghe72} and theoretical ZPE for triplet twisted C$_2$H$_4$~\cite{wang14}}\\
\multicolumn{11}{l}{$^\mathrm{d}$Exp.: 104.478$^\circ$~\cite{hoy79} (water) and 92.11$^\circ$~\cite{cook75} (hydrogen sulfide).}
\end{tabular}
\end{table*}

All three deformations exhibit a linear $\Sph$--$\Ecum$ relationship, with a very similar $\kappa$ for C--H-dissociation and C--C-dissociation and a smaller $\kappa$ for the torsion. The entropy increases according to the minimal orbital pairing approach relative to the low bond length limit: by $4\ln 2\approx 2.77$ for the C--H-dissociation and the torsion where only one bond is broken, and by $8\ln 2\approx 5.55$ for the C--C-dissociation. The observed value of $\kappa$ (0.129) for the C--H dissociation is very similar to that observed in hydrogen cyanide before (0.125), with a slightly smaller $b$ (0.122 instead of 0.139). For the C--C dissociation, $\kappa$ (0.126) is smaller than that found previously for the C--N-dissociation (0.135).

\begin{figure}
    \centering
    \includegraphics[width=\linewidth]{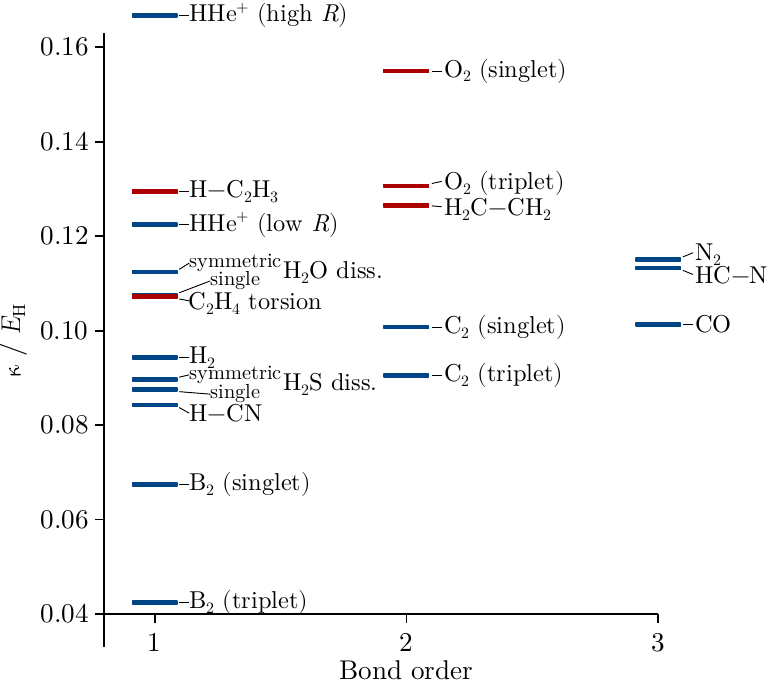}
    \caption{Distribution of best estimates for $\kappa$ values over all studied systems. Red data points: computed with STO-$n$G basis sets. Blue data points: computed with larger basis sets.}
    \label{fig:comparison}
\end{figure}

We conclude our investigation of the CC with a comparison of the best estimates for $\kappa$ for the systems studied in this work, which is presented in Fig.~\ref{fig:comparison}.

While some outliers lead to a high overall spread of $\kappa$ between 0.04 and 0.17, the majority of molecules, especially those with the chemically important bonds H--C, H--O, C--C, C--N, N--N, and C--O, lie within a more narrow range between 0.08 and 0.13. Given that the examples that yield $\kappa$ on the upper end of that range were computed with STO basis sets where $\kappa$ results in 20\%--40\% larger values than in cc-pVDZ, one can assume that range would further narrow to 0.08--0.115 in the basis set limit and include dioxygen as well. Furthermore, this would imply that a dissociation of a carbon--carbon double bond leads to a similar value of $\kappa$ in C$_2$ and ethylene. Carbon--hydrogen bonds, vary in $\kappa$ and depend on the chemical environment, which leads to different dissociation behaviors of HCN and C$_2$H$_4$.

Outside the range 0.08--0.115 lie two molecules that have been shown not to fulfill the CC well globally: the diboron molecule, where the fit has been performed only at a high bond distance where the slope reaches a minimum; and helium hydride, a molecule with a strongly polarized bond that is rather coordinative than covalent.

Within that range, there is a weak trend toward larger $\kappa$ values for higher-order bonds such as in N$_2$ which marks the upper limit. While covalent bonds with hydrogen have rather low values of $\kappa$ (such as in HCN or H$_2$S), water is an example where $\kappa$ is almost as large as in dinitrogen.

\subsection{Excited states}
Excited states are an interesting case for the application of the CC. Not only does the Rayleigh-Ritz variational principle not apply to their energy, but there is no constrained search functional to define the 2RDM based on the 1RDM. Still, the energy of excited states can be decomposed into a mean-field and a cumulant part via the 2RDM and an entropy can be calculated from the eigenvalues of the 1RDM. A further motivation for the study of excited states is our results on diatomics (see Sec.~\ref{sec:secondrow}) that show how dissociation curves can be shaped by crossings of the ground state and the excited state, leading to discontinuities in $\Ecum$ and $\Sph$.

\begin{figure*}
\includegraphics[width=.47\linewidth]{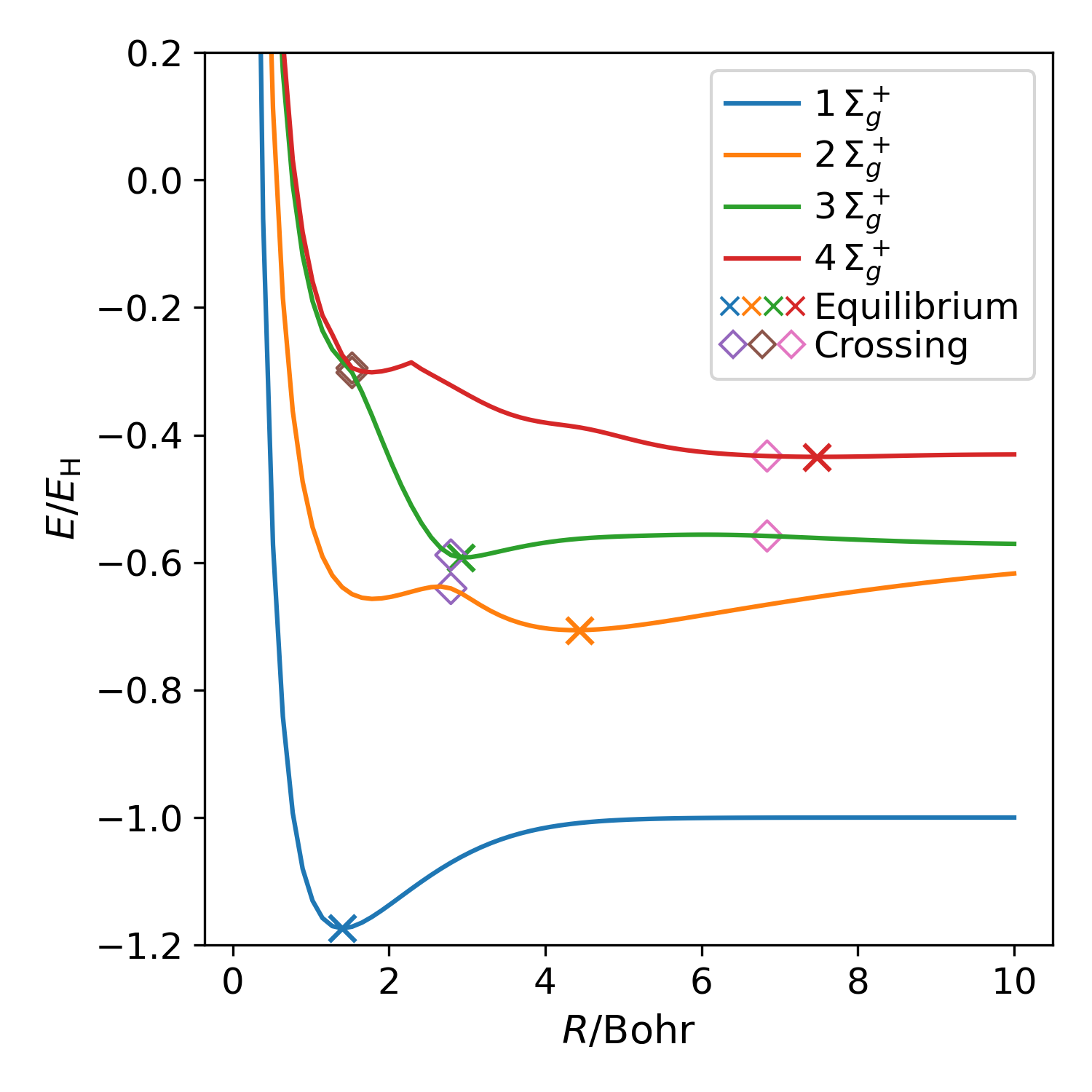}
\includegraphics[width=.47\linewidth]{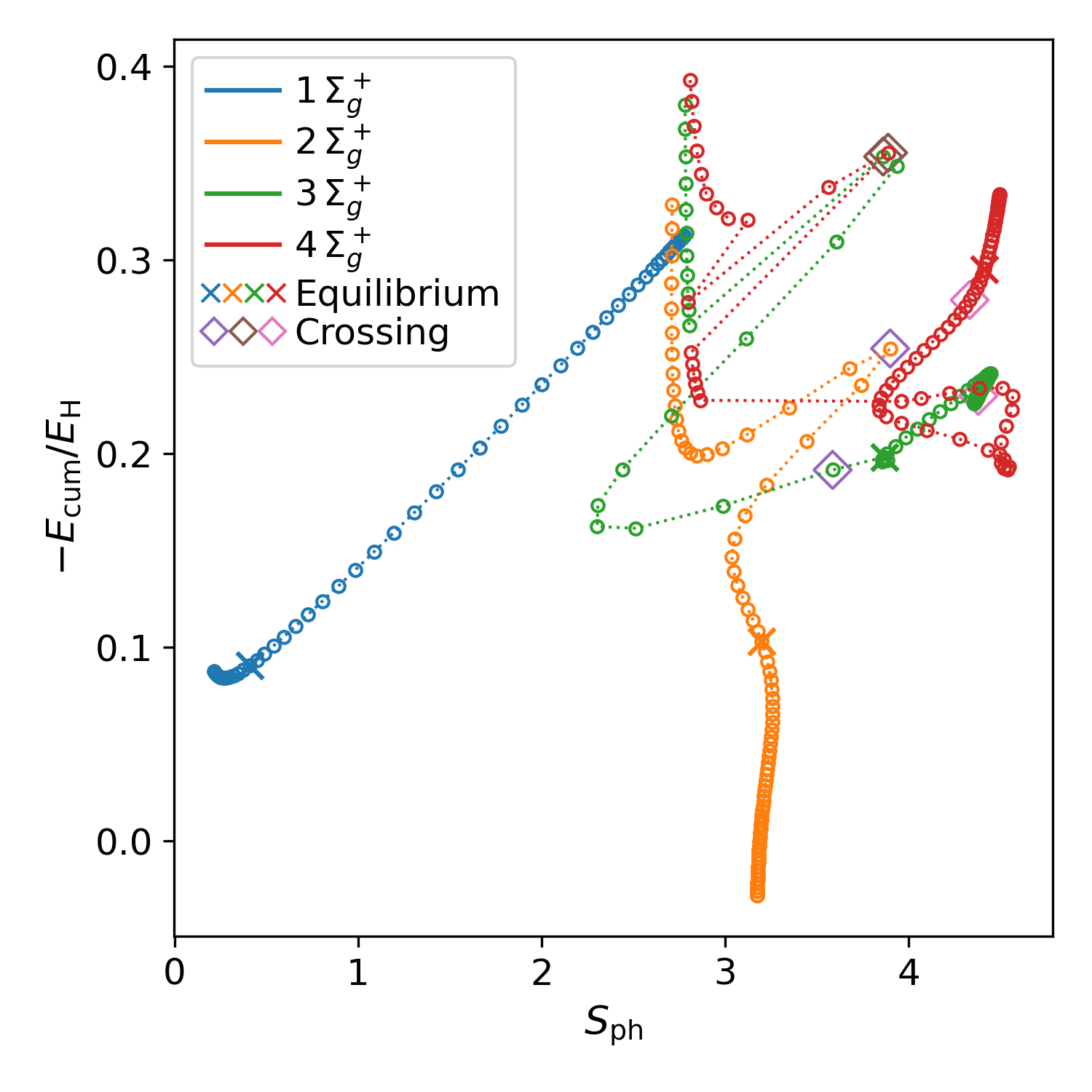}
\caption{Energy surfaces and cumulant energy $\Ecum$ vs particle-hole symmetrized entropy $\Sph$ for the lowest four $\Sigma_g^+$ states of dihydrogen{, evaluated with state-specific CASSCF(2,30)/cc-pV5Z calculations}.}\label{fig:h2sigmag}
\end{figure*}

As an example, we chose the dihydrogen molecule, on which we carried out state-specific CASSCF(2,30)/cc-pV5Z calculations. The large active space and basis set were chosen to achieve an equally good description of all excited states. The resulting energy surfaces (left panels) and $\Sph$--$\Ecum$ plots (right panels) for the four lowest-lying $^1\Sigma_g^+$ and $^1\Pi_u$  states are shown in Figs.~\ref{fig:h2sigmag} and \ref{fig:h2piu}, respectively. As a further example, data for $^1\Pi_g$ states are available in the supplementary material.

\begin{figure*}
\includegraphics[width=.47\linewidth]{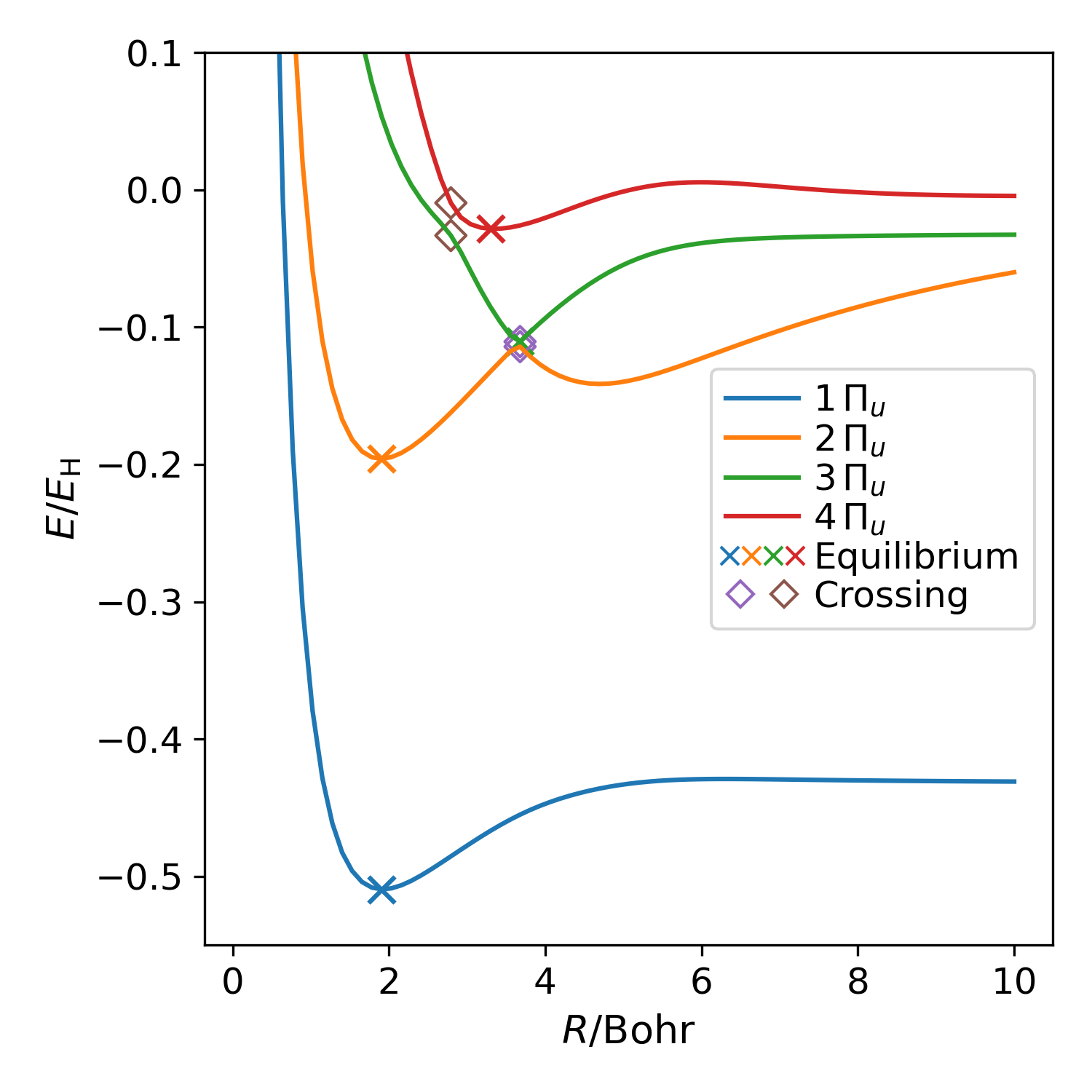}
\includegraphics[width=.47\linewidth]{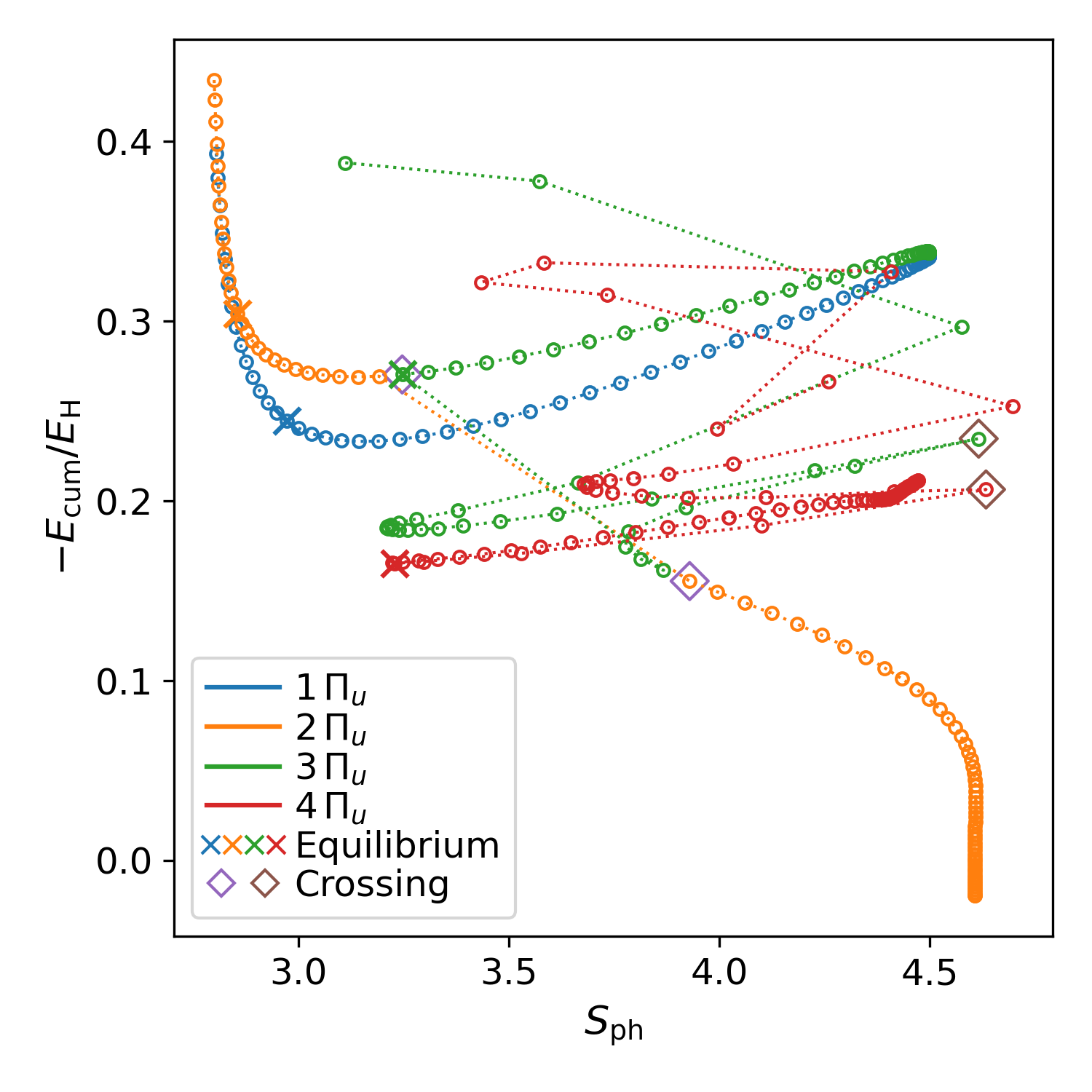}
\caption{Energy surfaces and cumulant energy $\Ecum$ vs particle-hole symmetrized entropy $\Sph$ for the lowest four $\Pi_u$ states of dihydrogen{, evaluated with state-specific CASSCF(2,30)/cc-pV5Z calculations}.}\label{fig:h2piu}
\end{figure*}

Among all three wave function symmetries, only the ground state is single-configurational at low bond lengths. It is also the only state that exhibits a globally linear relation between entropy and cumulant energy. The curves for the excited states do not exhibit a globally linear behavior whatsoever. While they have a few common features, the curves are dominated by discontinuities and often show vertical sections which implies that $\Ecum$ is not a functional of $\Sph$.

Toward the lower and in a few cases toward the upper limit of the entropies, the entropy change is minimal while the cumulant energy decreases during dissociation, such that the $\Sph$--$\Ecum$ curve appears vertical. For intermediate interatomic distances, the entropy typically varies strongly with $R$, and the cumulant energy changes approximately linearly. The slopes observed in these sections are typically lower than that of the ground state, and in some cases, even negative.

Many curves deviate from this general pattern and thus are further from linearity due to avoided crossings between states. In particular, one can distinguish these crossings into two types with different effects on the $\Sph$--$\Ecum$ relation. One case is crossings with low coupling strength, i.e., those where potential energy surfaces narrowly avoid each other, such as $3\Sigma_g^+$ and $4\Sigma_g^+$ or $2\Pi_g$ and $3\Pi_g$. Such narrowly avoided crossings do not lead to discontinuities or changes in slope or curvature of the $\Sph$--$\Ecum$ graph and the diabatic curves can be identified without problems. This is because states that mix the two electronic configurations significantly exist only in a very small range of geometries.

If states are subject to larger couplings, such as $2\Sigma_g^+$ and $3\Sigma_g^+$, $3\Pi_u$ or $4\Pi_u$, there is a much more pronounced effect on the entropy. In these regimes, mixed electronic states exist over a wider region of $R$. Depending on the exact nature of the two crossing states, this can lead to a strong increase or decrease in the entropy. This leads to a strong fluctuation of $\Ecum$ and $\Sph$ during the mixing of the electronic states, and they return to a smooth regime only after the crossing.

Overall, none of the excited states follow the CC. Even the $1\Pi_u$ state, the lowest-energy state in the $\Pi_u$ symmetry sector and significantly separated from the excited states, has a curved $\Sph$--$\Ecum$ relation and only reaches linearity well beyond the equilibrium distance. On the other hand, the $^1\Sigma_g^+$ state of dioxygen and the $^3\Sigma_g^-$ state of dicarbon that were studied in Sec.~\ref{sec:secondrow} do not represent the ground states of the respective spin states for these molecules, but were still found to agree well with the CC. These results point to special properties of the electronic ground state that enable the general applicability of the CC, but suggest that there might be an alternative form of the $\Sph$--$\Ecum$ relation for excited states that is compatible with the CC for some examples.

\section{Assessment of \MakeLowercase{i}-DMFT}
\label{sec:assess-iDMFT}

The positive results on the CC in the context of ground states for many of the investigated systems leaves us with an interesting prospect regarding its application as i-DMFT. We explore its abilities and limitations by comparing the results from i-DMFT for different physical quantities to a highly accurate CASSCF benchmark. For this, we implement the i-DMFT method as laid out in Ref.~\onlinecite{WangBaerends22-PRL} as a self-consistent field (SCF) routine in Python, building on the PySCF module. Details of the implementation underlying the calculations are presented in the supplementary material.

We re-assess the findings of \citet{WangBaerends22-PRL} for i-DMFT in dissociating H$_2$ and N$_2$. Furthermore, a fair assessment of the method's full potential necessitates investigations on more complex systems. Thus, i-DMFT is applied to the ethylene molecule, where the same three deformation processes are explored as in Sec.~\ref{sec:assess-cc} with details on the geometries given in the supplementary material. For each geometry, two i-DMFT calculations with different sets of parameters $\kappa$ and $b$ are carried out, which are listed in Table~\ref{tab:kappa_b_idmft}. The ``i-DMFT (CC)'' calculations employ the parameters obtained through linear fits to the $S_\mathrm{ph}$--$E_\mathrm{cum}$ curves in Sec.~\ref{sec:assess-cc}. The parameters in the ``i-DMFT (opt)'' calculations, on the other hand, are chosen to return the correct total energy at equilibrium and in full dissociation. For H$_2$ and N$_2$, optimized $\kappa$ and $b$ were adopted from Ref.~\citenum{WangBaerends22-PRL}, while for C$_2$H$_4$, we implemented an optimization scheme, details on which are given in the supplementary material.

\begin{table}[h!]
\caption{Results for $\kappa$ and $b$ for i-DMFT calculations in H$_2$, N$_2$ and C$_2$H$_4$.}\label{tab:kappa_b_idmft}
\resizebox{\columnwidth}{!}{%
\begin{tabular}{l|llc|llc|rr}\toprule
                     & \multicolumn{2}{c}{i-DMFT (CC)}               && \multicolumn{2}{c}{i-DMFT (opt)} && \multicolumn{2}{c}{Ratios} \\
                     & $\kappa$ / $E_\mathrm{H}$ & $b$ / $E_\mathrm{H}$ && $\kappa$ / $E_\mathrm{H}$ & $b$ / $E_\mathrm{H}$ && $\frac{\kappa^\mathrm{(opt)}}{\kappa^\mathrm{(CC)}}$ & $\frac{b^\mathrm{(opt)}}{b^\mathrm{(CC)}}$ \\\midrule
H$_2$                & 0.0967 & 0.0413 && 0.09468 & 0.02862 && 0.979 & 0.692 \\
N$_2$                & 0.119  & 0.226  && 0.1225 & 0.1439  && 1.029 & 0.636 \\
C$_2$H$_4$           &        &        &&          &           &&       &       \\
\quad C--C diss.      & 0.126  & 0.123  && 0.1052   & 0.1375    && 0.834 & 1.117 \\
\quad C--H diss.      & 0.129  & 0.122  && 0.1225   & 0.1112    && 0.949 & 0.911 \\
\quad CH$_2$ torsion & 0.107  & 0.165  && 0.1018   & 0.1411    && 0.951 & 0.855 \\\bottomrule
\end{tabular}
}
\end{table}

We find that the optimized values of $\kappa$ generally deviate by at most $5\%$ from the values from the linear fits. The only exception to this is the dissociation of the C--C bond in ethylene, where the energy-optimized $\kappa$ is significantly smaller than the parameter from the linear fit. As concluded from the results in Sec.~\ref{sec:assess-cc}, $\kappa$  primarily depends on static correlation. A good agreement between energy optimized and linearly fitted values of $\kappa$ thus indicates a correct treatment of static correlation in i-DMFT, as is typically seen for approximations within RDMFT. The deviation in the energy optimized values of $b$ is a lot larger. The interpretation of which, however, is not as clear as with $\kappa$. Formally, $b$ can be thought of as a linear shift, applied after an i-DMFT calculation has converged. By a sensible choice of an optimization scheme, it corrects for any systematic error in the energy curve. In parts, this is influenced by the treatment of the dynamic correlation in i-DMFT. In addition, from its definition in Sec.~\ref{sec:theory}, we understand that the i-DMFT functional is generally not variational for $\kappa$ and $b$ from the linear fit in the context of the CC, as
\begin{equation}
    \begin{split}
        E_\mathrm{gs}
        &=\trace[\hat{h}\gamma_\mathrm{gs}]+\F^\iDMFT(\gamma_\mathrm{gs})\\
        &\geq \inf_{\gamma}\left[\trace[\hat{h}\gamma]+\F^\iDMFT(\gamma)\right],
    \end{split}
\end{equation}
where the variational principle is applied over all 1RDMs. Thus, the true ground state energy will be underestimated. Both effects will be discussed thoroughly in the following Secs.~\ref{sec:idmft_gs_energy}--\ref{sec:idmft_ind_en_contr}.

\subsection{Ground state energy}\label{sec:idmft_gs_energy}

\begin{figure}
  \centering
  \includegraphics[width=.77\linewidth]{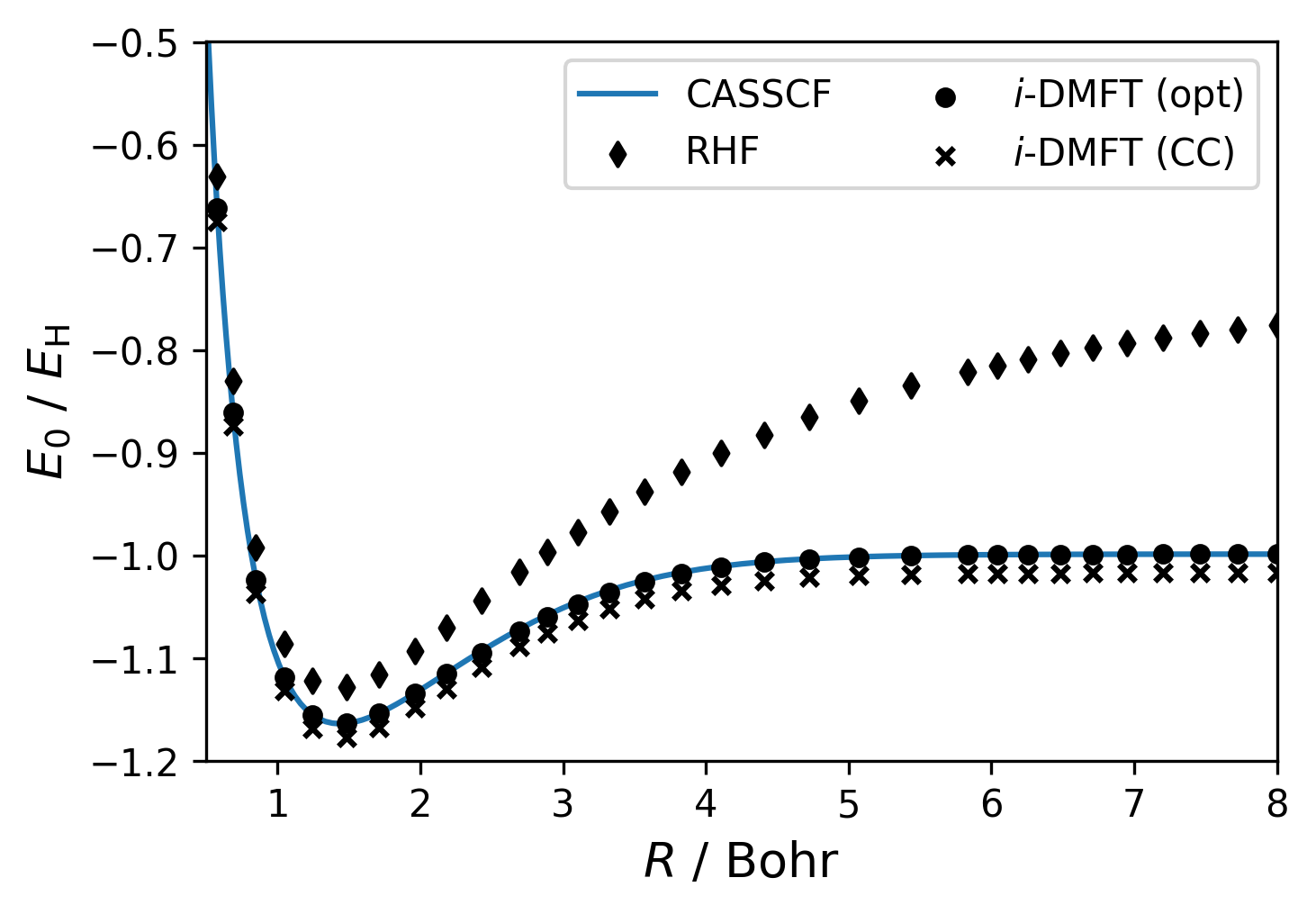}
  \caption{Total energy in different deformation processes of dihydrogen calculated with i-DMFT{/cc-pVDZ} and CASSCF{(2,10)/cc-pVDZ}. Two calculations with different parameters are presented for the i-DMFT method.}
  \label{fig:h2_idmft_e_tot}
\end{figure}

\begin{figure}
  \centering
  \includegraphics[width=.77\linewidth]{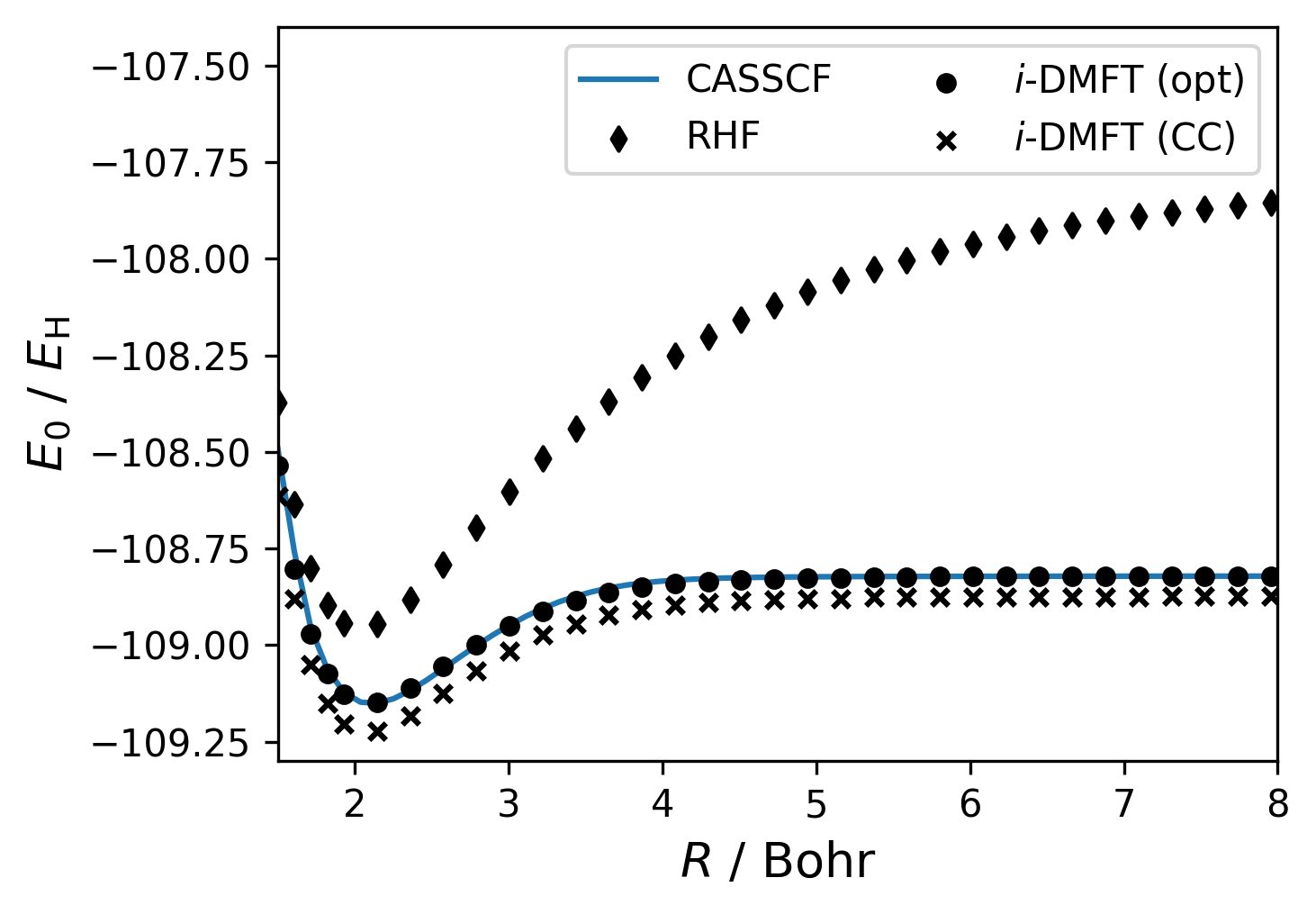}
  \caption{Total energy in different deformation processes of dinitrogen calculated with i-DMFT{/cc-pVDZ} and CASSCF{(6,24)/cc-pVDZ}. Two calculations with different parameters are presented for the i-DMFT method.}
  \label{fig:n2_idmft_e_tot}
\end{figure}

The central goal of i-DMFT is to obtain a good approximation of the ground state energy at mean-field cost over the course of a molecular deformation. Earlier publications \cite{WangBaerends22-PRL, WangWangSheng2021, Irimia2023-correlation, Irimia2023-open-shell, Liu2024-halogen, Hu2024-dispersion, Liu2025-halides} proved this to be successful in small molecules, including H$_2$ and N$_2$, with a sensible choice for the parameters $\kappa$ and $b$. Our results, using the cc-pVDZ basis set, confirm this, as shown in the energy curves in Figs.~\ref{fig:h2_idmft_e_tot} and \ref{fig:n2_idmft_e_tot} and Tabs.~\ref{tab:h2_idmft} and \ref{tab:n2_idmft}. {We note here that it was not possible to run CASSCF calculations for N$_2$ with all valence electrons active, while formally, all electrons and orbitals are active in i-DMFT. This discrepancy may lead to a different description of dynamic correlation between the two methods.}

\begin{table}
\caption{Fit results from the dissociation curves computed for hydrogen with i-DMFT in comparison to CASSCF$(n_\mathrm{e},n_\mathrm{orb})$ and RHF.}\label{tab:h2_idmft}
\begin{tabular}{lccclll}\toprule
Method & Basis & $n_\mathrm{e}$  & $n_\mathrm{orb}$ & $D_\mathrm{e}$/$E_\mathrm{H}$ & $r_\mathrm{e}$/bohr & $\tilde{\nu}$/cm$^{-1}$ \\\midrule
CASSCF                      & cc-pVDZ & 2 & 10 & 0.165 & 1.439 & 4429 \\
i-DMFT (opt)$^\mathrm{a}$ & cc-pVDZ & 2 & 10 & 0.165 & 1.436 & 4405 \\
i-DMFT (CC)$^\mathrm{a}$ & cc-pVDZ & 2 & 10 & 0.160 & 1.438 & 4385 \\
RHF                         & cc-pVDZ & 2 & 10 & 0.352 & 1.415 & 4629 \\\midrule
Exp.&&&&0.165&1.40&4401\\
&&&& \cite{liu09} & \cite{huber79} & \cite{huber79} \\\bottomrule
\multicolumn{7}{l}{$^\mathrm{a}$See Table~\ref{tab:kappa_b_idmft} for values of $\kappa$ and $b$.}\\
\end{tabular}
\end{table}

\begin{table}
\caption{Fit results from the dissociation curves computed for dinitrogen with i-DMFT in comparison to CASSCF$(n_\mathrm{e},n_\mathrm{orb})$ and RHF.}\label{tab:n2_idmft}
\begin{tabular}{lccclll}\toprule
Method & Basis & $n_\mathrm{e}$ & $n_\mathrm{orb}$ & $D_\mathrm{e}$/$E_\mathrm{H}$ & $r_\mathrm{e}$/bohr & $\tilde{\nu}$/cm$^{-1}$ \\\midrule
CASSCF                      & cc-pVDZ & 6  & 24 & 0.330 & 2.105 & 2368 \\
i-DMFT (opt)$^\mathrm{a}$ & cc-pVDZ & 14 & 28 & 0.330 & 2.083 & 2399 \\
i-DMFT (CC)$^\mathrm{a}$ & cc-pVDZ & 14 & 28 & 0.352 & 2.078 & 2437 \\
RHF                         & cc-pVDZ & 14 & 28 & 1.147 & 2.040 & 2749 \\\midrule
Exp.&&&&0.364&2.07&2359\\
&&&& \cite{lofthus77} & \cite{huber79} & \cite{irikura07} \\\bottomrule
\multicolumn{7}{l}{$^\mathrm{a}$See Table~\ref{tab:kappa_b_idmft} for values of $\kappa$ and $b$.}\\
\end{tabular}
\end{table}

On the other hand, they highlight that $\kappa$ and $b$ obtained through a linear fit of the $\Ecum$--$\Sph$ curve as in Sec.~\ref{sec:assess-cc} do not lead to a good approximation of the absolute value of the energy. The systematic underestimation demonstrates that the i-DMFT functional is not variational and, in particular, that Eq.~\eqref{eq:Ecum-inequality} is not fulfilled. The results for H$_2$ and N$_2$ from the restricted Hartree-Fock (RHF) method illustrate the shortcomings of the unaltered mean-field approach in strongly correlated systems: while a sensible approximation is found in the regime of weak correlation around equilibrium, RHF overestimates the total energy at large $R$ considerably. The unrestricted Hartree-Fock method can improve the performance in terms of energy, however, this comes at the cost of spin contamination in the minimizer.

\begin{figure*}
\includegraphics[width=\linewidth]{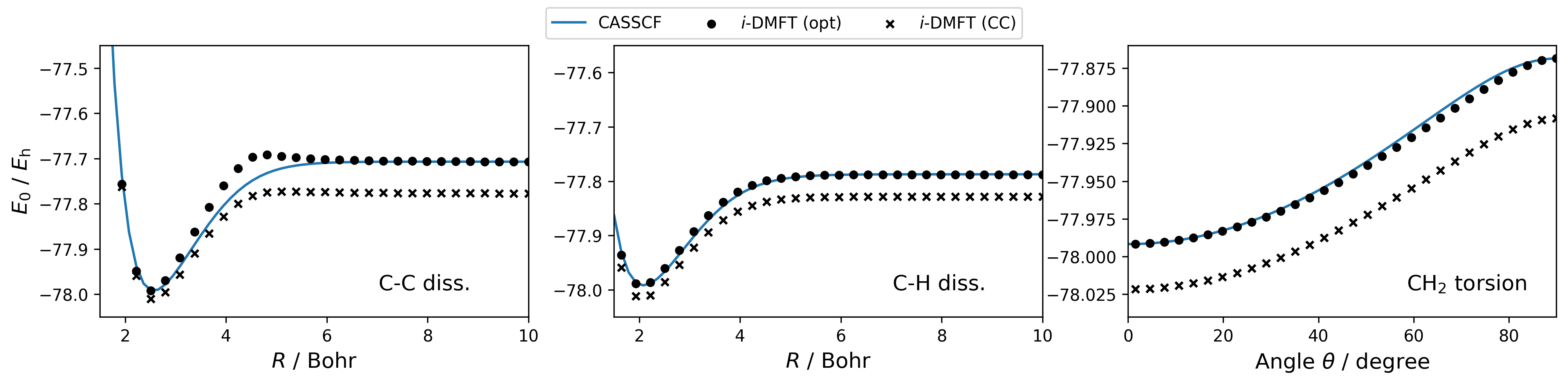}
\caption{Total energy in different deformation processes of ethylene calculated with i-DMFT{/STO-6G} and CASSCF{(12,12)/STO-6G}. Two calculations with different parameters are presented for the i-DMFT method.}
\label{fig:c2h4_idmft_e_tot}
\end{figure*}

In addition to the reproduction of these older results, we also carry out i-DMFT calculation for C$_2$H$_4$ in the STO-6G basis set, which are displayed in Fig.~\ref{fig:c2h4_idmft_e_tot} and whose numerical results are presented in Table~\ref{tab:c2h4_idmft}.
The agreement between i-DMFT (opt) and CASSCF results is good for dissociation of the single C--H bond and the torsion of the CH$_2$ groups. For the dissociation of the C--C bond the overall performance of the i-DMFT (opt) calculations is significantly worse for intermediate values of $R$. i-DMFT (opt) produces an energetic barrier around $4.5$\,bohr, which is not present in the benchmark. The formation of the barrier is caused by the small value of $\kappa$ in the energy optimized calculation, which only accounts for $83.4\%$ of the parameter from the linear fit of the $S_\mathrm{ph}$--$E_\mathrm{cum}$ curve. Thus, for $R$ slightly larger than the equilibrium distance, the mean-field part of the functional dominates, which leads to a significant overestimation of the total energy in that regime. Once the total energy value is saturated, the construction scheme for $\kappa$ and $b$ ensures good agreement with the reference.

\subsection{Entropy and NONs}

\begin{table*}
\caption{Fit results from the dissociation curves computed for ethylene with i-DMFT in comparison to CASSCF$(n_\mathrm{e},n_\mathrm{orb})$ and RHF.}\label{tab:c2h4_idmft}
\begin{tabular}{llccclll}\toprule
Deformation & Method & Basis & $n_\mathrm{e}$ & $n_\mathrm{orb}$ & $D_\mathrm{e}$/$E_\mathrm{H}$ & $r_\mathrm{e}$/bohr & $\tilde{\nu}$/cm$^{-1}$ \\\midrule
C-C diss.      & CASSCF                      & STO-6G & 12 & 12 & 0.285 & 2.565 & 1571 \\
               & i-DMFT (opt)$^\mathrm{a}$ & STO-6G & 16 & 14 & 0.285 & 2.515 & 1693 \\
               & i-DMFT (CC)$^\mathrm{a}$ & STO-6G & 16 & 14 & 0.233 & 2.555 & 1569 \\
& Exp. & & & & 0.286~\cite{chase82,furtenbacher06,lerberghe72} & 2.530~\cite{herzberg66} & 1623~\cite{lerberghe72}$^\mathrm{b}$\\
C-H diss.      & CASSCF                      & STO-6G & 12 & 12 & 0.204 & 2.084 & 3385 \\
               & i-DMFT (opt)$^\mathrm{a}$ & STO-6G & 16 & 14 & 0.204 & 2.051 & 3519 \\
               & i-DMFT (CC)$^\mathrm{a}$ & STO-6G & 16 & 14 & 0.187 & 2.056 & 3470 \\
Exp. & & & & & 0.186~\cite{atct,cccbdb,lerberghe72} & 2.052~\cite{herzberg66} & 3026~\cite{lerberghe72}$^\mathrm{b}$\\\midrule
               &                             &        &    &    &       & $\theta_\mathrm{e}$/$^\circ$ & $k_2$/$E_\mathrm{H}$Rad$^{-2}$ \\\midrule
CH$_2$ torsion & CASSCF                      & STO-6G & 12 & 12 & 0.123$^\mathrm{c}$ & 0 & 0.150 \\
               & i-DMFT (opt)$^\mathrm{a}$ & STO-6G & 16 & 14 & 0.123$^\mathrm{c}$ & 0 & 0.143 \\
               & i-DMFT (CC)$^\mathrm{a}$ & STO-6G & 16 & 14 & 0.113$^\mathrm{c}$ & 0 & 0.137 \\\bottomrule
\multicolumn{8}{l}{$^\mathrm{a}$See Table~\ref{tab:kappa_b_idmft} for values of $\kappa$ and $b$.}\\
\multicolumn{8}{l}{$^\mathrm{b}$Vibrational mode resembling the symmetric stretching.}\\
\multicolumn{8}{l}{$^\mathrm{c}$Exp.: 0.119~$E_\mathrm{H}$ from thermal rotation barrier~\cite{douglas55}, ZPE for C$_2$H$_4$~\cite{lerberghe72} and theoretical }\\
\multicolumn{8}{l}{\hspace{26.5pt} ZPE for triplet twisted C$_2$H$_4$~\cite{wang14}}\\
\end{tabular}
\end{table*}

\begin{figure*}
\centering
\begin{minipage}{.45\linewidth}
  \centering
  \includegraphics[width=.77\linewidth]{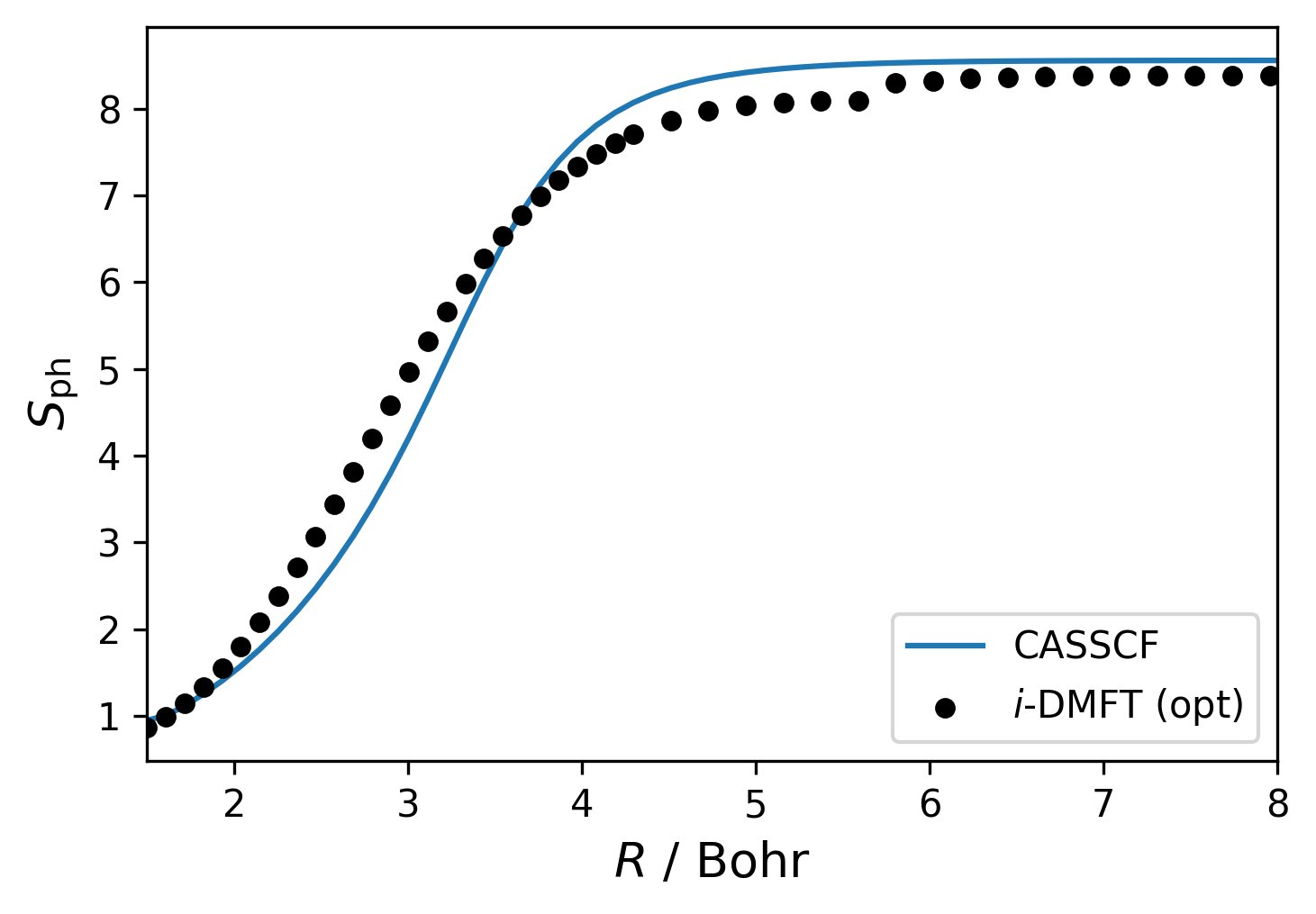}
  \caption{Entropy along the dissociation of dinitrogen from the i-DMFT{/cc-pVDZ} (opt) calculation and the CASSCF{(6,24)/cc-pVDZ} benchmark.}
  \label{fig:n2_idmft_entropy}
\end{minipage}%
\hspace{20pt}
\begin{minipage}{.45\textwidth}
  \centering
  \includegraphics[width=.77\linewidth]{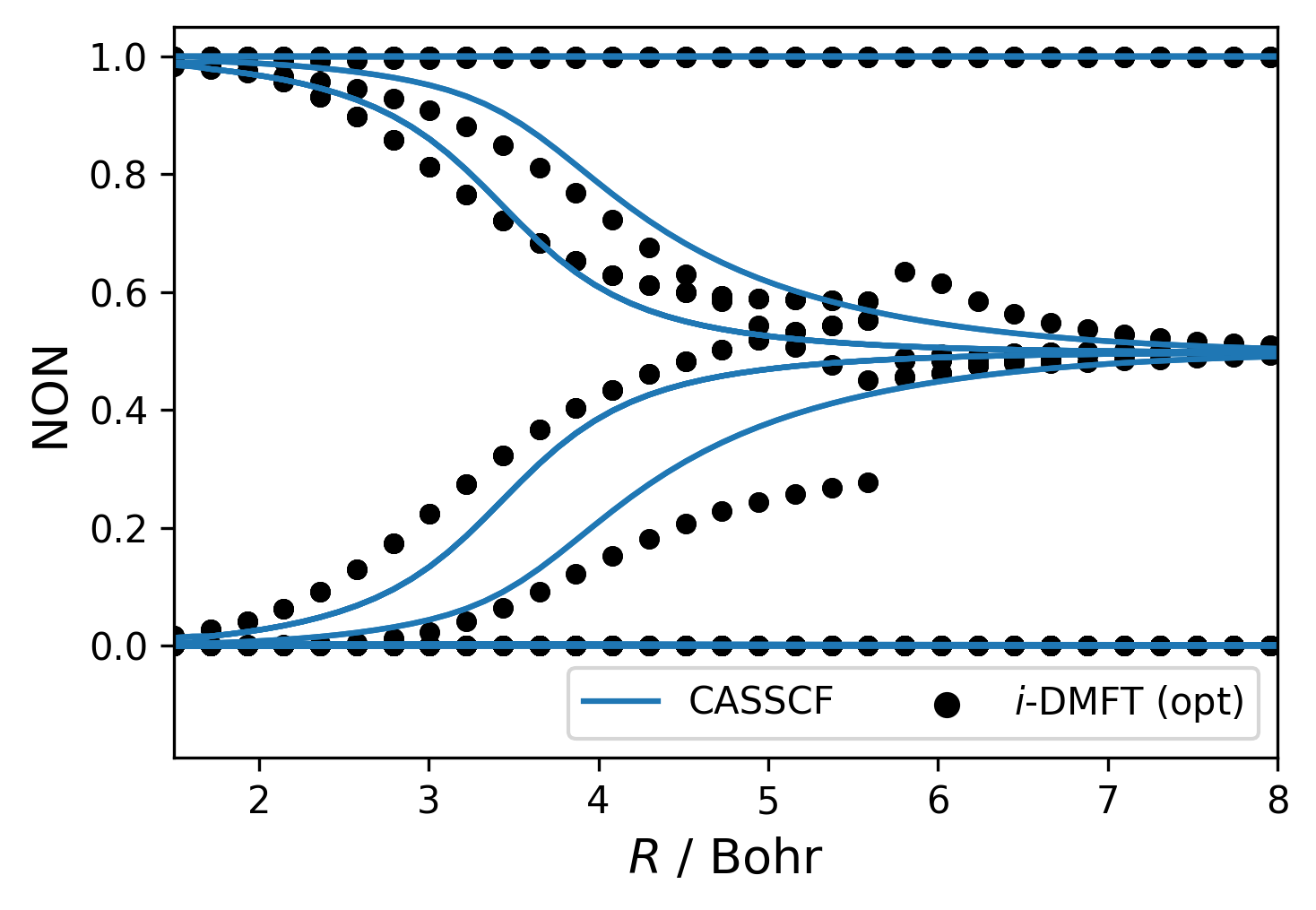}
  \caption{Evolution of the NONs along the dissociation of dinitrogen from the i-DMFT{/cc-pVDZ} (opt) calculation and the CASSCF{(6,24)/cc-pVDZ} benchmark.}
  \label{fig:n2_idmft_noon}
\end{minipage}
\end{figure*}

\begin{figure*}
\includegraphics[width=\linewidth]{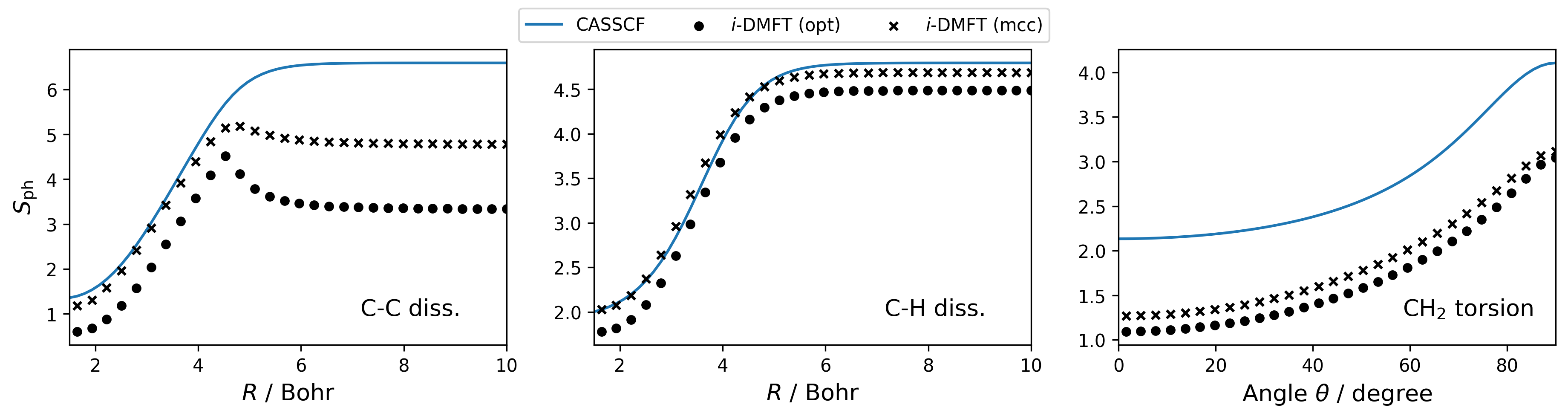}
\caption{Entropy along different deformation processes of ethylene calculated with i-DMFT{/STO-6G} and CASSCF{(12,12)/STO-6G}.}
\label{fig:c2h4_idmft_entropy}
\end{figure*}

\begin{figure*}
\includegraphics[width=\linewidth]{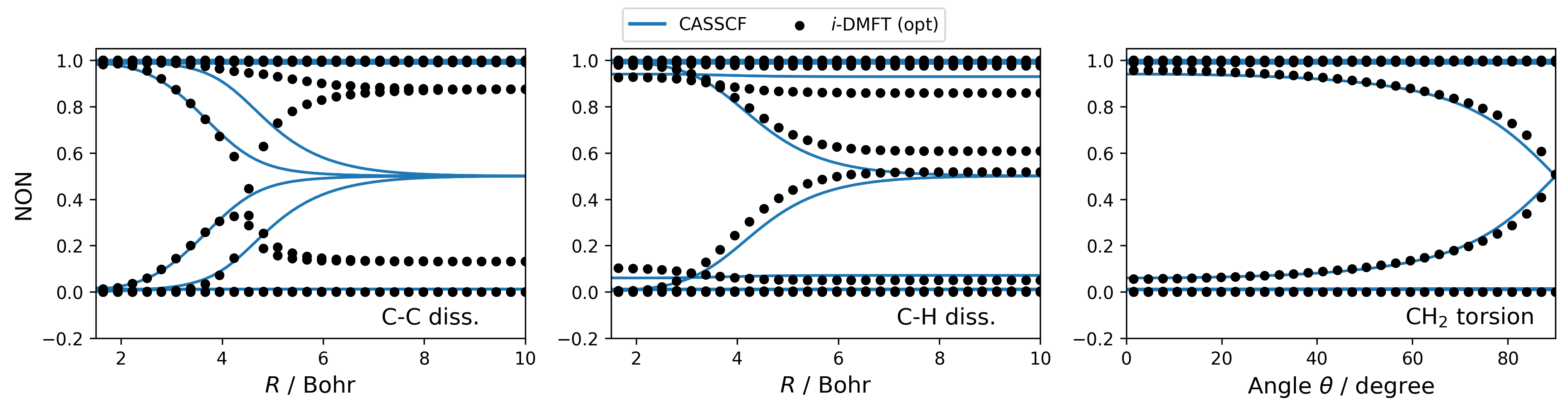}
\caption{Evolution of the NONs for different deformation processes of ethylene, calculated with i-DMFT{/STO-6G} (opt) and CASSCF{(12,12)/STO-6G}.}
\label{fig:c2h4_idmft_noon}
\end{figure*}

As already shown by Ref.~\onlinecite{WangBaerends22-PRL}, an i-DMFT calculation in H$_2$ with energy-optimized values of $\kappa$ and $b$ returns a good approximation for the evolution of the NONs and $S_\mathrm{ph}$. Again, this is confirmed by our findings, which are presented in the supplementary material. We perform the same analysis in Figs.~\ref{fig:n2_idmft_entropy} and~\ref{fig:n2_idmft_noon} for N$_2$ and in Figs.~\ref{fig:c2h4_idmft_entropy} and~\ref{fig:c2h4_idmft_noon} for C$_2$H$_4$.


In N$_2$, $S_\mathrm{ph}$ from i-DMFT (opt) aligns well with the CASSCF benchmark at large dissociation distances and around the equilibrium. For intermediate values of $R$ the agreement is slightly worse, until a sudden increase improves the performance for $R>5.5$\,bohr. This is also reflected in the evolution of the NONs, where a sharp change in behavior is observed around $R=5.5$\,bohr, with particularly poor agreement for smaller values of $R$.

In ethylene, the agreement between entropy from the i-DMFT (opt) and benchmark calculations is poor. While the qualitative behavior of $S_\mathrm{ph}$ is reproduced well for the dissociation of a single C--H bond and the torsion of the CH$_2$ groups, a constant offset is present. A similar behavior is also found with the C--C bond dissociation for $R$ up to $4.5$~bohr. Dissociation beyond that distance sees a decrease in $S_\mathrm{ph}$ from i-DMFT (opt) which results in a particularly bad agreement.

The error in entropy for the different deformations in C$_2$H$_4$ results from the evolution of the NONs. Good agreement with the benchmark is found for the CH$_2$ torsion. The offset in the entropy is explained by a constant error in the NONs close to occupations of $0$ and $1$. In the dissociation of a single C--H bond, the overall qualitative behavior is reproduced well, however, with wrong occupations at small values of $R$ and in the dissociation limit. As expected from the investigation on the entropy, poor agreement between NONs from i-DMFT and the benchmark is found for the dissociation of the C--C bond at $R>4.5$\,bohr. It is observed, however, that the behavior of NONs in the left panel of Fig.~\ref{fig:c2h4_idmft_noon} resembles that for dissociation of the C--C bond in the excited ethylene~\cite{Cheung1979}, which might provide an explanation for the large deviations in the total energy over the course of dissociation in Fig.~\ref{fig:c2h4_idmft_e_tot}.

Independent of the overall performance in terms of NONs, in all systems we find significant errors in the occupation of orbitals which are either nearly empty or fully occupied. Namely, the occupations are too close to $0$ and $1$, when compared to the CAS benchmark. In N$_2$ this effect is largely suppressed as our benchmark has only six electrons active in 24 orbitals, which fixes the occupation of core orbitals to exactly $1$. In C$_2$H$_4$, however, the full extent of this deficiency of i-DMFT is highlighted for the torsion of the CH$_2$ groups: despite capturing the overall trend in the behavior of the NONs well, a significant offset is observed in the entropy. This indicates a poor treatment of dynamic correlation by i-DMFT.

All of these results taken together suggest that the minimizer in i-DMFT features considerable deviations from benchmark results also beyond the NONs themselves. This is substantiated by the subsequent discussion.

\subsection{1RDM}

While a good performance of i-DMFT in terms of the total energy for many of the investigated systems supports its applicability, its success must also be assessed by its ability to reproduce the correct 1RDM, the central variable in RDMFT. This requires a quantitative measure of the deviation between the i-DMFT 1RDM $\gamma_{\iDMFT}$ and the CASSCF reference $\gamma_\mathrm{CAS}$. To this end, we employ the trace distance,
\begin{equation}\label{eq:tr_dist}
    \begin{split}
        {D}(\gamma_{\iDMFT},\gamma_\mathrm{CAS})=
        \frac{1}{2n_\mathrm{e}}\trace\left[\sqrt{\left(\gamma_{\iDMFT}-\gamma_\mathrm{CAS}\right)^2}\right],
    \end{split}
\end{equation}
where both 1RDMs are expressed in the same orthonormal basis. The number of electrons in the system is denoted by $n_\mathrm{e}$. In order to allow for comparison of the result between systems with different numbers of electrons, the normalization of $1/2n_\mathrm{e}$ ensures that $0 \leq D \leq 1$.

\begin{figure*}
\centering
\begin{minipage}{.45\linewidth}
  \centering
  \includegraphics[width=.77\linewidth]{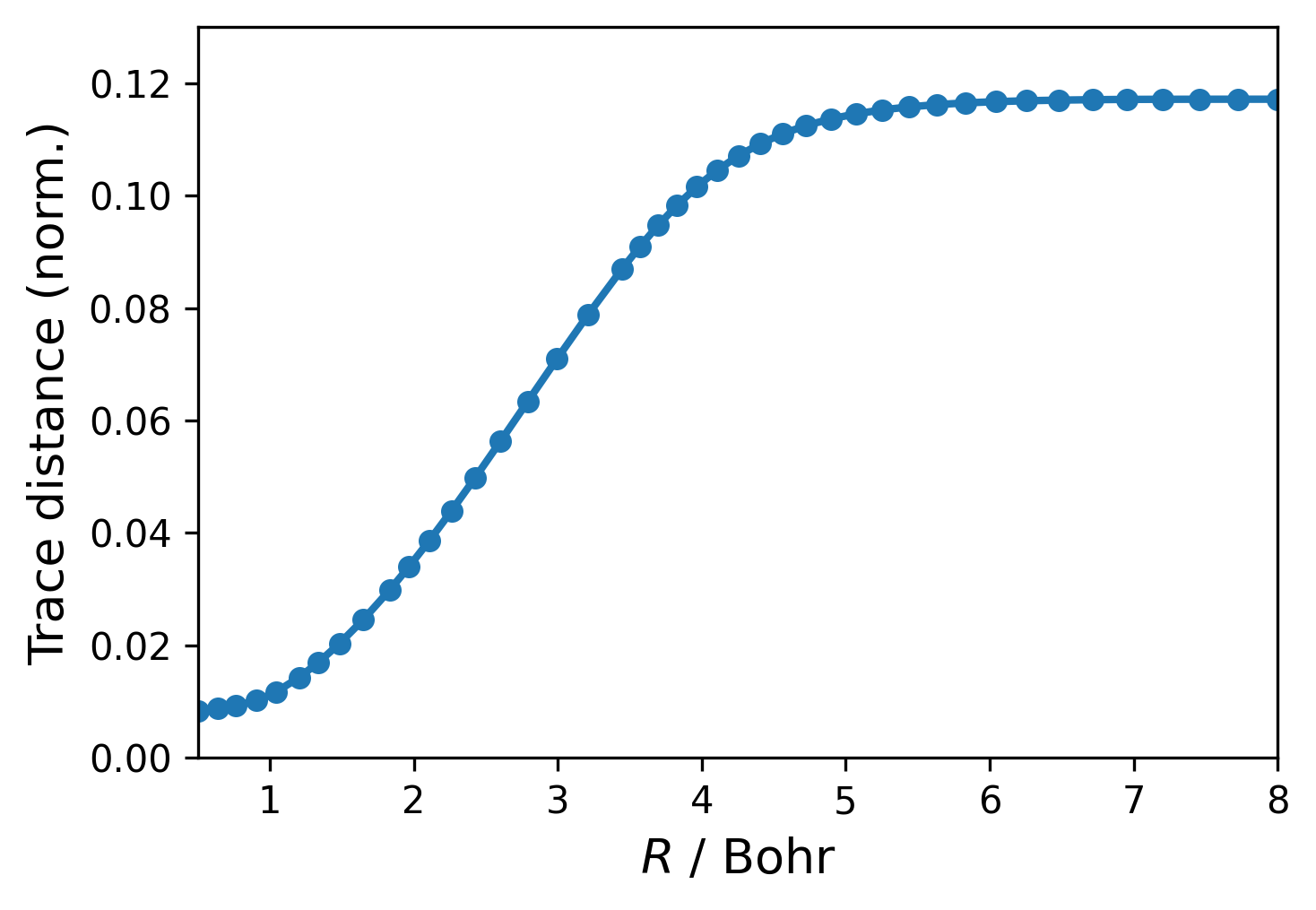}
  \caption{Normalized error in the 1RDM computed with i-DMFT{/cc-pVDZ} along the dissociation of hydrogen with respect to the CASSCF{(2,10)/cc-pVDZ} result.}
  \label{fig:h2_idmft_err_1rdm}
\end{minipage}%
\hspace{20pt}
\begin{minipage}{.45\textwidth}
  \centering
  \includegraphics[width=.77\linewidth]{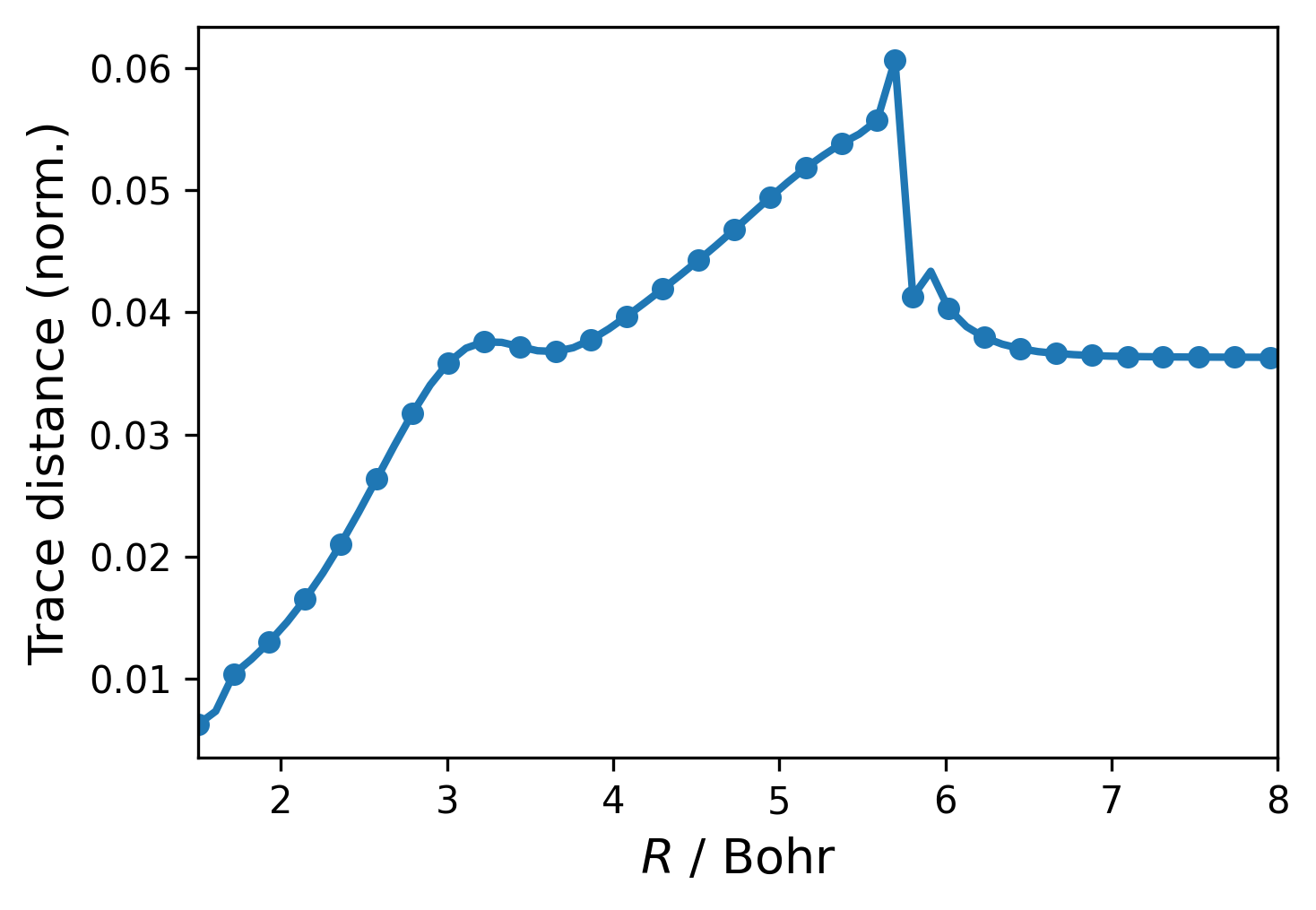}
  \caption{Normalized error in the 1RDM computed with i-DMFT{/cc-pVDZ} along the dissociation of dinitrogen with respect to the CASSCF{(6,24)/cc-pVDZ} result.}
  \label{fig:n2_idmft_err_1rdm}
\end{minipage}
\end{figure*}
\begin{figure*}
\includegraphics[width=\linewidth]{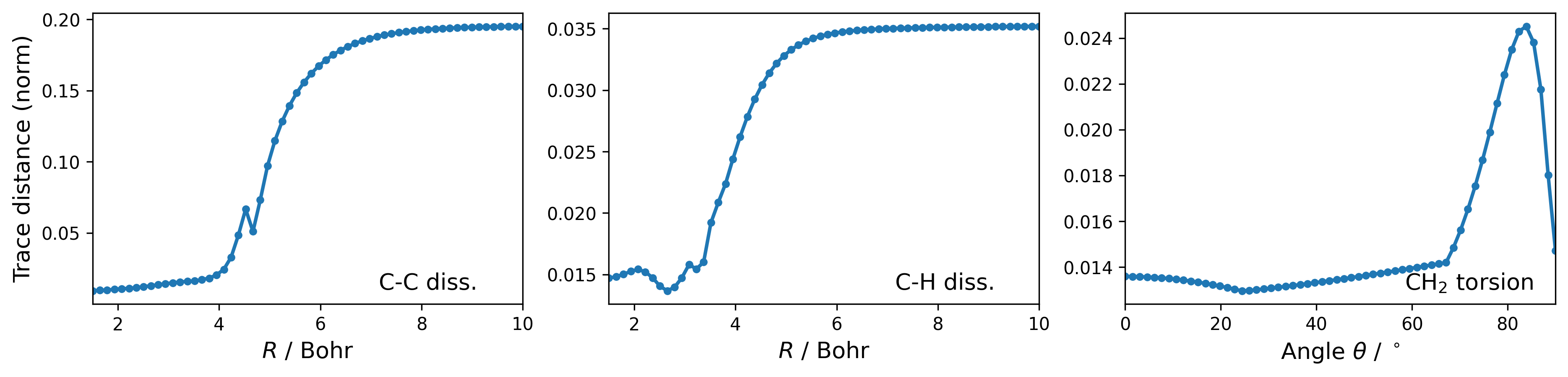}
\caption{Normalized error in the 1RDM computed with i-DMFT{/STO-6G} in different deformation processes with respect to the CASSCF{(12,12)/STO-6G} result.}
\label{fig:c2h4_idmft_err_1rdm}
\end{figure*}

As of Figs.~\ref{fig:h2_idmft_err_1rdm}--\ref{fig:c2h4_idmft_err_1rdm}, the error of the 1RDM is comparatively minor for geometries close to the equilibrium in all systems investigated. Generally, the error then increases significantly over the course of the deformation. A unique behavior is observed for the torsion of the CH$_2$ groups in ethylene (right panel of Fig.~\ref{fig:c2h4_idmft_err_1rdm}): the error remains relatively small for a large interval of torsion angles until it spikes at around $\theta=80^\circ$, only to drop quickly right before the final geometry at $\theta=90^\circ$. Sharp changes in the behavior of the NONs translate into abrupt shifts in $D$. This is demonstrated most clearly for N$_2$, where $D$ rises uncharacteristically from around $R=3.5$\,bohr, until it drops within a very small interval of $R$ around $5.5$\,bohr back to a smaller value, which persists until full dissociation. A similar artifact is also present in the left panel of Fig.~\ref{fig:c2h4_idmft_err_1rdm} around $R=4.5$\,bohr, where the dissociation of the C--C bond in ethylene is investigated.
Furthermore, recent investigations of the fermionic exchange force~\cite{CS19ExForce} reveal that the behavior of the i-DMFT functional near minimal and maximal occupation numbers ($n \to 0$ and $n \to 1$) is qualitatively incorrect. In particular, its derivative with respect to the occupation numbers should diverge as $-\partial F/\partial n \sim 1/\sqrt{n}$ for $n \to 0$ and $1/\sqrt{1-n}$ for $n \to 1$, whereas the corresponding gradient in i-DMFT exhibits only a weaker, logarithmic divergence and thus fails to reproduce the correct exchange force.
When comparing the values of the trace distance to the NONs from Figs.~\ref{fig:n2_idmft_noon} and~\ref{fig:c2h4_idmft_noon} (as well as the curve for H$_2$ in the supplementary material), it is found that the error in the 1RDM is not always explained by a deviation in the NONs. While most pronounced at large dissociation distances in H$_2$ and N$_2$, this effect is observed for most geometries in all systems. This indicates errors in the 1RDMs beyond the NONs, i.e., a poor approximation of the NOs.

\begin{figure*}
\includegraphics[width=.7\linewidth]{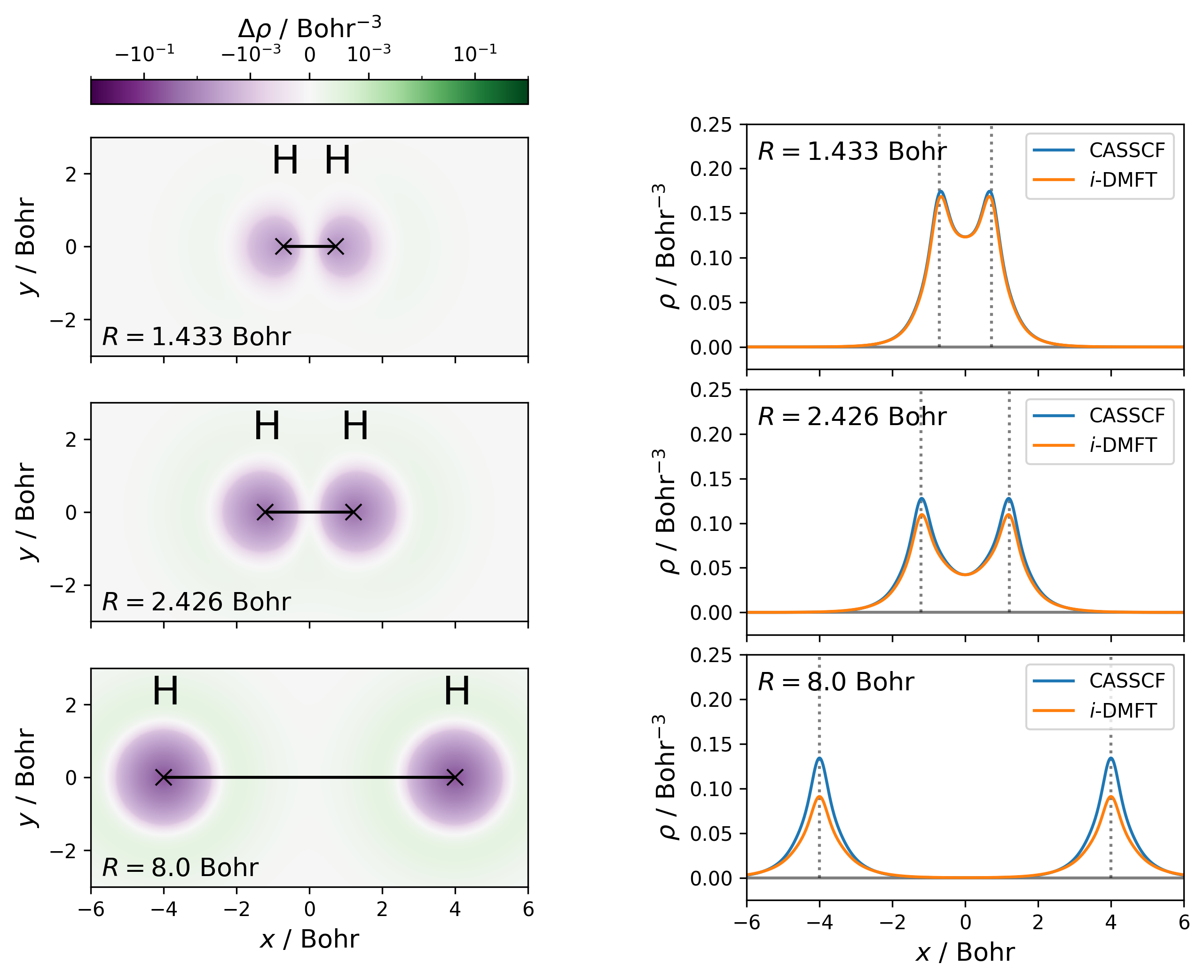}
\caption{Left column: normalized error in charge densities of hydrogen between i-DMFT{/cc-pVDZ} (opt) and a CASSCF{(2,10)/cc-pVDZ} calculation in a plane containing the H--H bond. Right column: normalized charge density along molecular axis computed with i-DMFT{/cc-pVDZ} (opt) in comparison to the CASSCF{(2,10)/cc-pVDZ} result. The top, middle, and bottom panels show results for different internuclear distances.}
\label{fig:h2_idmft_err_charge_density}
\end{figure*}

Inevitably, this structural error translates further into errors in one-electron observables. This includes the electronic charge density
\begin{equation}
    \rho(\rr)=\trace[\hat c_\rr^\dagger\hat c_\rr \gamma],
\end{equation}
with $\hat c_\rr^{(\dagger)}$ being the annihilation (creation) operator of the position state corresponding to $\rr$. Our results in Fig.~\ref{fig:h2_idmft_err_charge_density} show that the i-DMFT results --- even for an optimal choice of $\kappa$ and $b$ in terms of the total energy in the given basis set --- fail to capture the correct charge density: the panels in the left column display the difference $\Delta{\rho}$ between normalized electronic charge densities ${\rho}_{\iDMFT}$ and ${\rho}_\mathrm{CASSCF}$ from i-DMFT and CASSCF calculations,
\begin{equation}\label{eq:charge_density_difference}
    \Delta{\rho}(\rr)={\rho}_{\iDMFT}(\rr)-{\rho}_\mathrm{CASSCF}(\rr),
\end{equation}
in a plane that contains the molecular axis. Here, we are using normalized densities such that
\begin{equation}
    \int_{\mathbb{R}^3}{\rho}(\rr)\rmd^3r=1.
\end{equation}
The deviation is visualized by use of a colormap in a symmetric logarithmic format: values scale logarithmically in the intervals $[-10^0,-10^{-3}]$ and $[10^{-3}, 10^0]$ and linearly between $-10^{-3}$ and $10^{-3}$.
The right column of Fig.~\ref{fig:h2_idmft_err_charge_density} displays the normalized charge densities from i-DMFT and CASSCF along the molecular axis. For every value of the dissociation variable $R$, i-DMFT systematically underestimates the concentration of charge in the vicinity of the nucleus. Instead, the charge is more diffuse, similar to the result expected from an RHF calculation. By distributing the charge over a larger spatial volume, the effect of dynamic correlation is reduced, as was already seen in the context of our analysis of the NONs. This substantiates even further our previous assumption that the stark mismatch in the values of $b$ from the optimization procedure and the linear fit in Sec.~\ref{sec:assess-cc} is rooted --- at least in parts --- in a poor description of dynamic correlation, which extends the error in the charge distribution to the value of the total energy.

\subsection{Individual energy contributions}\label{sec:idmft_ind_en_contr}

The error in the charge density naturally leads to errors in the one-particle contributions to the total energy. Here, we decompose the energy into kinetic energy, potential energy, and further interaction energy. The errors in kinetic and potential energy, $\Delta E_\mathrm{kin}$ and $\Delta E_\mathrm{pot}$, are obtained as
\begin{equation}
    \begin{split}
        &\Delta E_\mathrm{kin}=\trace\left[\hat T(\gamma_\iDMFT-\gamma_\mathrm{CAS})\right], \\
        &\Delta E_\mathrm{pot}=\trace\left[\hat V(\gamma_\iDMFT-\gamma_\mathrm{CAS})\right], \\
    \end{split}
\end{equation}
with $\hat T,\hat V$ being the kinetic and potential energy operators, respectively. The error in the interaction energy, on the other hand, is calculated as
\begin{equation}
    \Delta E_\mathrm{int}=\F^\iDMFT(\gamma_\iDMFT)-\trace\left[\hat{W}\Gamma_\mathrm{CAS}\right],
\end{equation}
with $\Gamma_\mathrm{CAS}$ being the full ground-state density matrix from the CASSCF calculation, while $\F^\iDMFT$ is taken from Eq.~\eqref{func}, where energy optimized values of $\kappa$ and $b$ are to be used.

\begin{figure*}
\centering
\begin{minipage}{.45\linewidth}
  \centering
  \includegraphics[width=.77\linewidth]{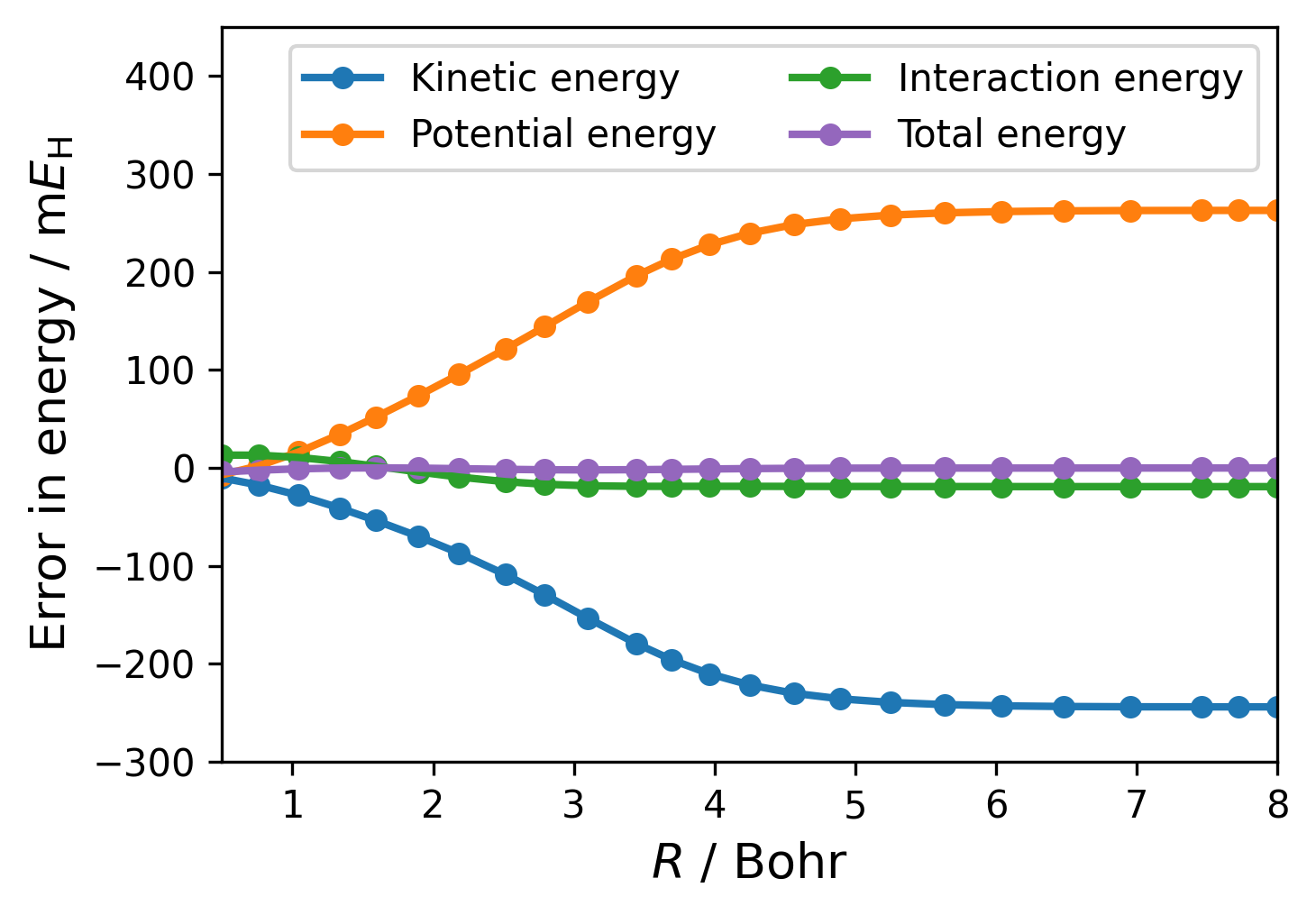}
  \caption{Error in individual i-DMFT{/cc-pVDZ} (opt) energy contributions along the dissociation of hydrogen compared to CASSCF{(2,10)/cc-pVDZ} results.}
  \label{fig:h2_idmft_energy_contr}
\end{minipage}%
\hspace{20pt}
\begin{minipage}{.45\textwidth}
  \centering
  \includegraphics[width=.77\linewidth]{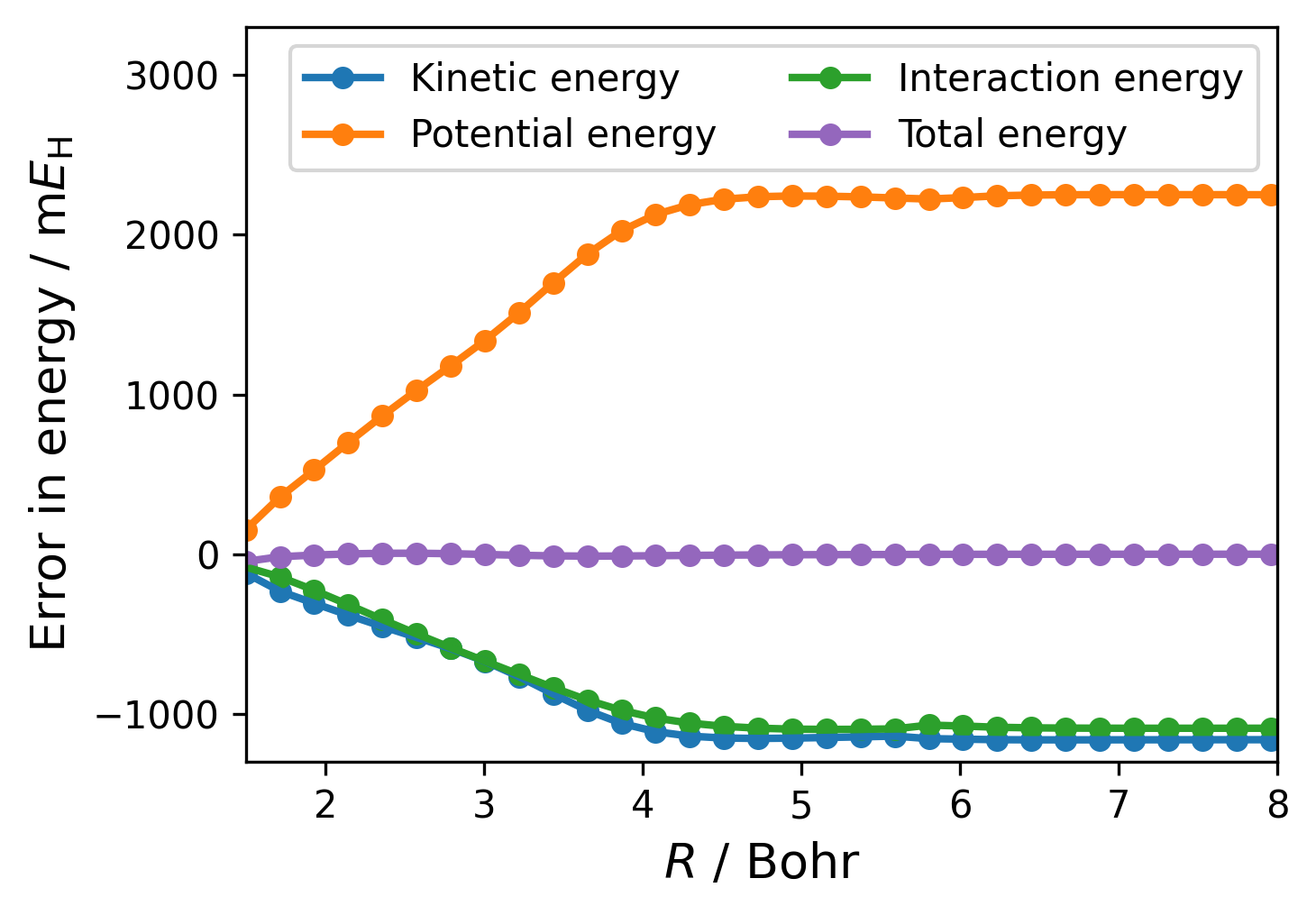}
  \caption{Error in individual i-DMFT{/cc-pVDZ} (opt) energy contributions along the dissociation of dinitrogen compared to CASSCF{(6,24)/cc-pVDZ} results.}
  \label{fig:n2_idmft_energy_contr}
\end{minipage}
\end{figure*}

\begin{figure*}
\includegraphics[width=\linewidth]{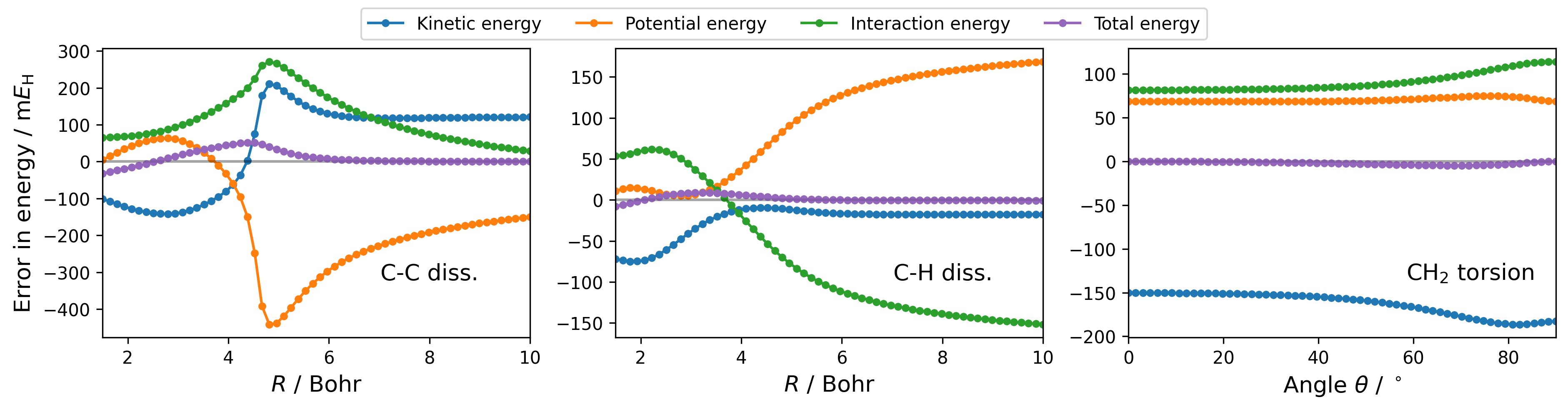}
\caption{Error in individual i-DMFT{/STO-6G} (opt) energy contributions along different dissociation processes of ethylene compared to CASSCF{(12,12)/STO-6G} results.}
\label{fig:c2h4_idmft_energy_contr}
\end{figure*}

The results for the different molecules are displayed in Figs.~\ref{fig:h2_idmft_energy_contr}--\ref{fig:c2h4_idmft_energy_contr}.
Both in H$_2$ (Fig.~\ref{fig:h2_idmft_energy_contr}) and N$_2$ (Fig.~\ref{fig:h2_idmft_energy_contr}), the potential energy stemming from electron--nucleus interaction is underestimated. This aligns well with our previous observation on the overly diffusive character of the charge distribution in i-DMFT. In H$_2$ the error is on the order of $100\,\mathrm{m}E_\mathrm{h}$, while in N$_2$ it is on the order of $1\,E_\mathrm{h}$. By construction of the optimization procedure for $\kappa$ and $b$, this error is compensated by the errors in the kinetic energy of the electrons and the electron--electron interaction energy, which leads to only minor deviations in the total energy.

In C$_2$H$_4$ (Fig.~\ref{fig:c2h4_idmft_energy_contr}), the behavior of the energy contributions is more complicated. For all three deformations we find errors in the individual energies on the order of $100$\,m$E_\mathrm{H}$. For the dissociation of the C--C bond, the errors are most pronounced in the vicinity of $R=4.5$\,bohr, which aligns with our previous observations. The dissociation of the C--H bond shows an underestimation of the magnitude of the nucleus--electron potential energy at large dissociation distances ($R>4$\,bohr) similar to our findings in H$_2$ and N$_2$. The torsion of the CH$_2$ groups returns errors in the individual contributions that remain almost constant over the deformation. An increase is noted around $\theta=80^\circ$, which matches the deformation where a large value for the trace distance $D$ is observed in Fig.~\ref{fig:c2h4_idmft_err_1rdm}.

From these findings, it is clear that the success of i-DMFT in approximating the total energy does not extend to an accurate description of the individual energy contributions. Critically, even the one-particle observables in the Hamiltonian show significant deviations from their CASSCF benchmark. This further underlines the crucial point that the 1RDM -- as the central variable within RDMFT -- is described poorly.

\section{Conclusions and outlook}\label{sec:concl}

Motivated by the promise of i-DMFT to capture static correlation at mean-field cost, we have provided a comprehensive assessment of the validity of the correlation/Collins' conjecture (CC) and the applicability of i-DMFT on a variety of chemical systems. For this, we analyzed and compared the behavior of entropy and cumulant energy in symmetric (B$_2$, C$_2$, N$_2$, O$_2$) and unsymmetric (CO) second-row diatomics including different spin states. We also reported on the first ever investigation of the CC for polyatomic systems with different types of bonds (HCN, C$_2$H$_4$) and studied a bond breaking process that is induced by a bond torsion instead of a dissociation; for a third-row element by directly comparing H$_2$S with H$_2$O; and for excited states. 

Our investigations showed that the molecules with single-configurational equilibrium ground states fulfill the CC during dissociation to a very good degree, even though none of the examples showed numerically exact linearity. This even holds during torsion of the ethylene molecule, where one of the bonds between the carbon atoms is broken without a dissociation. 
Bond compression ultimately leads to a departure from linearity, but fulfills the CC in the chemically meaningful range of 15--25\% that is relevant for vibrational dynamics.

The exceptions that we found are the helium hydride cation and some diatomics with open-shell equilibrium electron configurations such as B$_2$ and C$_2$.
This is related to two effects: First, the natural occupation number pattern during their dissociation cannot be predicted based on a minimal orbital pairing ansatz.
Second, there are parts of the ground-state potential energy surface with avoided crossings with excited states, for which we could show that the CC is not satisfied.

This allows us to formulate three criteria that indicate that the CC can be expected to remain valid during bond breaking:
\begin{tcolorbox}
\begin{enumerate}[wide,leftmargin=*]
    \item The electron reorganization is merely a redistribution within pairs of orbitals, e.g., bonding/antibonding orbitals.
    \item The bond breaking is not completely heterolytic.
    \item The electronic states involved are electronic ground states and there are no avoided crossings.
\end{enumerate}
\end{tcolorbox}
If one of these criteria is not valid, one can expect deviations from the conjecture in the shape of a jump or transition between different linear segments, or a nonzero curvature over the entire dissociation. {These criteria already impose significant limitations on ab initio methods derived from the CC, in particular because the confirmation of the presence or absence of shifts in electron distribution requires higher-level calculations.}

For the values of $\kappa$ and $b$, we found a consistent basis set and active space dependence: Minimal single-zeta basis sets systematically overestimate $\kappa$ and underestimate $b$, but already cc-pVDZ basis sets estimate $\kappa$ reliably when including all strongly correlated electrons, in the active space. On the other hand, $b$ depends on dynamic correlation and only converges with large active spaces. As originally conjectured, the parameters vary from molecule to molecule, but for many common bonds, $\kappa$ lies in a narrow range between 0.08 and 0.115~$E_\mathrm{H}$, and results on C--C double bonds give reason to assume that there is little variation of $\kappa$ for the same kind of bond in different chemical environments.
Furthermore, the results on helium hydride seem to indicate that larger slopes are to be expected in coordinative bonds. 

On the method level, we investigated the performance of i-DMFT for a range of scenarios. We applied it to H$_2$ and N$_2$, examples that were already considered before~\cite{WangBaerends22-PRL}, and on three different deformation processes in C$_2$H$_4$. The results demonstrate that i-DMFT based on parameters $\kappa$ and $b$ obtained from a linear fit to CASSCF results does not result in a good approximation of the total energy, but in a systematic underestimation. 
This implies that the functional is not variational over the domain of $N$-representable 1RDMs even when the parameters are fitted to the exact entropies and cumulant energies. Instead, 1RDMs exist that yield a lower energy than the exact ground-state energy.
Better overall agreement is obtained by determining $\kappa$ and $b$ from a fit to the total energy at equilibrium and in the dissociation limit.

Comparison of NONs from i-DMFT and CASSCF reveals that i-DMFT yields occupation numbers too close to the boundary, implying a poor description of dynamic correlation. Because of the close relation of these NONs and the entropy, this directly affects the total energy. {Furthermore, this error in the occupation numbers typically also implies a failure in the computation of one-electron properties computed from the 1RDM.}
This effect is present in all investigated cases, but is most pronounced in ethylene where a large number of orbitals contribute to dynamic correlation. While the fitting of $b$ to the total energy can partially include correlation as a constant offset, deviations in the curvature of energy curves remain.

The need for an ad-hoc optimization of $\kappa$ and $b$ calls the conceptual ties of i-DMFT to the CC into question. The special role that is attributed to the entropy $\Sph$ because of its success in the context of the CC might be less justified when using it in a correlation functional. Generally, any concave function of the NONs will also lead to fractional occupation numbers, which is vital for describing (strongly) correlated systems, and different choices have been discussed before~\cite{Ziesche1995,Goedecker1998,Liu2025-halides}.

Across all systems studied, i-DMFT generally fails to accurately describe properties beyond energy surfaces. Notably, even optimized $\kappa$ and $b$ values yield poor absolute energies in the C--C bond dissociation in ethylene, where the NON behavior suggests that i-DMFT captures an excited state rather than the ground state.
Possible modifications to i-DMFT to achieve a wider applicability will be discussed in a future publication.

Information-theoretical methods already help to design more efficient wave function methods~\cite{stein16,bensberg23,ding23} or density-functional approaches~\cite{FZX25}.
Our findings here underscore their crucial role also in RDMFT, a promising alternative to computationally expensive wave-function methods and Kohn–Sham DFT. Through the use of entropy in the framework of i-DMFT, RDMFT can be realized as a method with mean-field cost.
A critical next step would be a rigorous demarcation of a range of model chemistries where the CC can be proven to hold exactly, pointing out areas of applicability more clearly. This can also help to extend the method beyond the current scope, for example by including higher-order dependencies on the entropy and correcting for the error regarding the boundary force. This way, the correct decay behavior for the NONs~\cite{Giesbertz2013,Cioslowski2019,CiS21,sobolev2021a} linked to dynamic correlation can be recovered and the description of the total electron density improved. Finally, the method would greatly benefit from further first-principles determination approaches for $\kappa$ and $b$, ideally embedded into a self-consistent scheme, and in the best case replacing the current empirical and system-dependent scheme. Such advances will help to unlock RDMFT's full potential and drive the next paradigm shift in computational quantum chemistry.

\section*{Supplementary Material}
The supplementary material for this paper is attached to this document. It contains details regarding the computation, implementation, and fitting/optimization procedures and specifications of the molecular geometries, as well as additional numerical data that is not presented in the main paper.

\begin{acknowledgments}
We thank J.~Cioslowski and P.~Knowles for their helpful discussions and E.~J.~Baerends and J.~Wang for an informative correspondence. We acknowledge the financial support from the German Research Foundation (Grant SCHI 1476/1-1) (L.D., J.L., M.P., C.S.), the International Max Planck Research School for Quantum Science and Technology (IMPRS-QST) (J.L.), and the ERC-2021-STG under grant agreement No.~101041487 REGAL (M.P.). The project is part of the Munich Quantum Valley, which is supported by the Bavarian state government with funds from the Hightech Agenda Bayern Plus.
\end{acknowledgments}

\bibliography{refs}

\clearpage

\newcount\p
\p=1
\loop
    \includepdf[pages=\the\p]{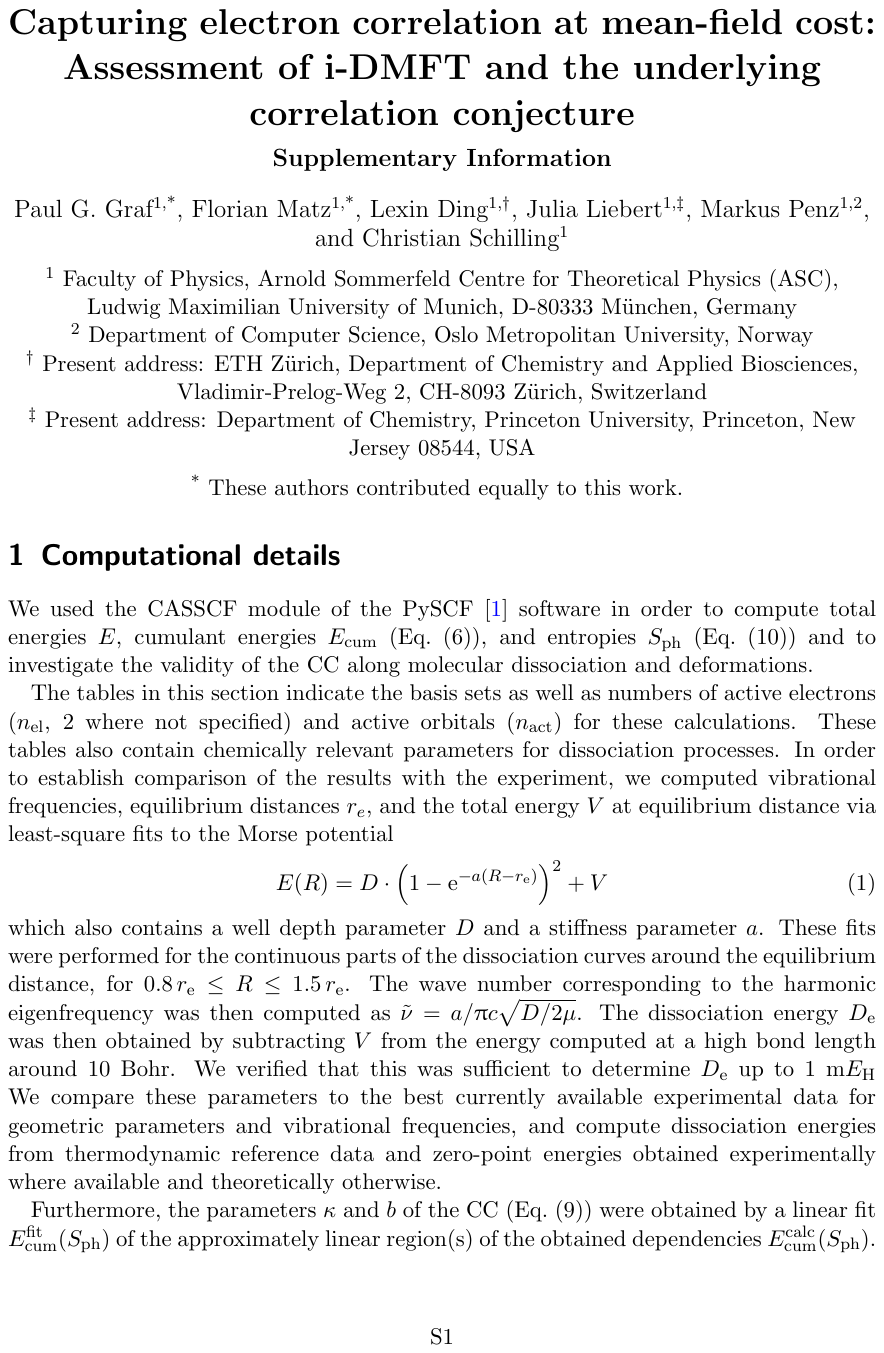}
    \clearpage
    \advance\p by 1
\ifnum\p<10
\repeat

\end{document}